\documentclass[12pt,a4paper]{article}%
\usepackage{amssymb}
\usepackage{amsfonts}
\usepackage{amsmath}
\usepackage{graphicx}
\usepackage[singlespacing]{setspace}
\usepackage{geometry}
\usepackage{url}%
\providecommand{\U}[1]{\protect\rule{.1in}{.1in}}

\begin{document}

\title{\textbf{HOPF\ BIFURCATION\ FROM }\\\textbf{NEW-KEYNESIAN\ TAYLOR\ RULE }\\\textbf{TO\ RAMSEY\ OPTIMAL\ POLICY}}
\author{Jean-Bernard Chatelain\thanks{Paris School of Economics, Universit\'{e} Paris
I Pantheon Sorbonne, 48 Boulevard Jourdan 75014 Paris. Email:
jean-bernard.chatelain@univ-paris1.fr} and Kirsten Ralf\thanks{ESCE
International Business School, Inseec U Research Center, 10 rue Sextius
Michel, 75015 Paris, Email: Kirsten.Ralf@esce.fr.}\bigskip
\and Post-Print: \textit{Macroeconomic Dynamics}, 17th January 2020, Pages 1-33.
\and https://doi.org/10.1017/S1365100519001032}
\maketitle

\begin{abstract}
This paper compares different implementations of monetary policy in a
new-Keynesian setting. We can show that a shift from Ramsey optimal policy
under short-term commitment (based on a negative feedback mechanism) to a
Taylor rule (based on a positive feedback mechanism) corresponds to a Hopf
bifurcation with opposite policy advice and a change of the dynamic
properties. This bifurcation occurs because of the \emph{ad hoc} assumption
that interest rate is a forward-looking variable when policy targets
(inflation and output gap) are forward-looking variables in the new-Keynesian theory.

\textbf{JEL\ classification numbers}: C61, C62, E43, E44, E47, E52, E58.

\newpage

\end{abstract}

\section{INTRODUCTION}

The question of how to conduct monetary policy is one of the most discussed
questions in macroeconomic theory. In this paper, we will compare Ramsey
optimal policy which is based on a negative feedback mechanism to a Taylor
rule which is based on a positive feedback mechanism. Both settings are
analyzed in the same New-Keynesian benchmark model with households'
intertemporal optimization of consumption and a New-Keynesian Phillips curve.

On the one hand, negative feedback is the core mechanism for stabilizing
dynamic systems with optimal control (Astr\"{o}m and Kumar (2014), Chatelain
and Ralf (2019b)). On the other hand, determinacy, as required by rational
expectations theory, is often associated with a positive feedback mechanism in
the New-Keynesian setting. Barnett and Chen (2015) and Barnett and Duzhak
(2008, 2010), therefore, emphasize the importance of bifurcations in the
reference New-Keynesian macroeconomic model (Gali (2015)). The Taylor rule
parameters (the response of the interest rate to inflation or to the output
gap) are bifurcation parameters. A small change of their values may lead to
big changes of the dynamic path, from stability to instability and conversely.

The aim of this paper is to clarify these issues further, extending the
bifurcation results of Barnett and Duzhak (2008 and 2010) on the closed
economy New-Keynesian model. We use Wonham's (1967) pole placement theorem to
locate linear feedback rule parameters for two basic types of policy:
New-Keynesian Taylor rule and Ramsey optimal policy.

For Ramsey optimal policy under quasi-commitment, the policy targets
(inflation and output gap) are forward-looking. Optimal behavior implies that
the policy instruments (the interest rate and its lag) are optimally
predetermined, using initial transversality conditions. This leads to negative
feedback Taylor rule parameters since higher inflation would lead the Fed to
set interest rates in a way that produces \emph{lower} future inflation. This
mechanism is in the spirit of Keynesian stabilization theory.

The private sector agents have a unique optimal anchor of forward-looking
output gap and inflation on the unique optimal initial values of the
\emph{observable} central bank funds rate and its lag. With Ramsey optimal
policy, negative feedback Taylor rule parameters are compatible with
determinacy, which corresponds to the uniqueness of the path of
forward-looking variables.

In the New-Keynesian model, the only equation which is not derived from
optimal intertemporal rational choice is the Taylor rule. Hence, it is the
only equation where the researcher writing down the model can
\emph{arbitrarily} decide if the variable whose behavior is governed by the
Taylor rule equation, the central bank's funds rate, is predetermined or
forward-looking. This \emph{ad hoc} theoretical assumption is an input in
order to apply Blanchard and Kahn's (1980) determinacy condition. If the
central bank's funds rate is assumed to be forward-looking for forward-looking
inflation and output gap, positive feedback Taylor rule parameters lead to
determinacy, with only one possible value of inflation suiTable for the
central bank. As Cochrane (2011) states: \textquotedblleft\textit{In
New-Keynesian models, higher inflation leads the Fed to set interest rates in
a way that produces even higher future inflation. For only one value of
inflation today will inflation fail to explode.}"

Inflation and output gap are forward-looking variables that need to be
anchored on predetermined variables. If the central bank's funds rate is a
forward-looking variable, it cannot be an anchor. Hence, the private sector
agents are assumed to anchor inflation and output gap on \emph{non-observable}
variables, which are two \emph{ad hoc }autoregressive predetermined shocks: a
cost-push shock and a productivity shock.

We can show that the shift from Ramsey optimal policy, with two stable
eigenvalues, to a Taylor rule, with two unstable complex conjugate
eigenvalues, corresponds to a Hopf bifurcation which is only driven by the
choice of a theoretical assumption by researchers on central bank's funds rate
as being optimally predetermined variables (Ramsey optimal policy) or \emph{ad
hoc} forward-looking variable (Taylor rule).

The \emph{ad hoc} assumption that policy instrument should be forward-looking
instead of backward-looking when the policy targets are forward-looking is the
origin of the bifurcation of the dynamic systems when assuming New-Keynesian
Taylor rule in place of Ramsey optimal policy.

Hence, our policy recommendation would be to use Ramsey optimal policy with a
reduced form negative feedback Taylor rule. For the monetary transmission
channel, empirical work may confirm the relevance of a delayed cost-of-capital
channel of monetary policy where interest rate is negatively correlated with
future output as opposed to an intertemporal substitution channel where the
interest rate is positively correlated with future output gap. If this is the
case, a negative feedback Taylor rule parameter of the response of interest
rate to output gap is positive. Else, it would be negative.

The rest of the paper is organized as follows: In Section 2, the model and its
dynamic properties are presented, including Barnett and Duzhak's (2008, 2010)
bifurcations frontier. In Section 3, Ramsey optimal policy and a Taylor rule
are compared. Section 4 shows that a shift from Ramsey optimal policy to a
New-Keynesian Taylor rule corresponds to a Hopf bifurcation. In Section 5, we
analyze the consequences when changing the assumptions on the exogenous
shocks. Section 6 concludes.

\section{BIFURCATIONS\ IN\ THE\ NEW-KEYNESIAN\ MODEL}

\subsection{The New-Keynesian Model}

The New-Keynesian private sector's four-equations model is written with all
variables as log-deviations of an equilibrium (Gali (2015)). In the
representative household's intertemporal substitution consumption Euler
equation, the expected future output gap $E_{t}x_{t+1}$ is \emph{positively}
correlated with the real rate of interest, equal to the nominal rate $i_{t}$
minus \emph{expected} inflation $E_{t}\pi_{t+1}$. The intertemporal elasticity
of substitution (IES) $\gamma=1/\sigma$ is a measure of the responsiveness of
the growth rate of consumption to the interest rate, usually considered to be
smaller than one. It is the inverse of $\sigma$, the relative fluctuation
aversion or the relative degree of resistance to intertemporal substitution of
consumption, which measures the strength of the preference for smoothing
consumption over time, usually considered to be larger than one.%

\begin{equation}
x_{t}=E_{t}x_{t+1}-\gamma\left(  i_{t}-E_{t}\pi_{t+1}\right)  +z_{t}\text{
with }\gamma>0.
\end{equation}

A non-controllable exogenous stationary and predetermined variable $z_{t}$ is
autoregressive of order one ($0<\left\vert \rho_{z,x}\right\vert <1$) where
$\varepsilon_{z,t}$ are zero-mean, normally, independently and identically
distributed additive disturbances. The initial value of the predetermined
forcing variable is given.%

\begin{equation}
\text{ }z_{t}=\rho_{z}z_{t-1}+\varepsilon_{z,t}\text{, }z_{0}\text{ given.}%
\end{equation}

In the New-Keynesian Phillips curve, expected inflation $E_{t}\pi_{t+1}$ is
negatively correlated with the current output gap $x_{t}$ with a sensitivity
$-\kappa<0$.
\[
\pi_{t}=\beta E_{t}\pi_{t+1}+\kappa x_{t}+u_{t}\text{ with }1>\beta>0\text{
and }\kappa>0.
\]

\emph{Sign restrictions} are such that parameters $\gamma,\beta,\kappa$ are
all strictly positive. A non-controllable exogenous stationary and
predetermined variable cost-push shock $u_{t}$ is\ autoregressive of order one
( $0<\left\vert \rho_{u}\right\vert <1$) where $\varepsilon_{u,t}$ are
zero-mean, normally, independently and identically distributed additive
disturbances. The initial value of the predetermined cost-push shock $u_{0}$
is given.%

\begin{equation}
\text{ }u_{t}=\rho_{u}u_{t-1}+\varepsilon_{u,t}\text{ , }u_{0}\text{ given.}%
\end{equation}

Firstly, the controllable dynamics of the New-Keynesian model with its
feedback Taylor rule can be written as follows:%

\[
\left(
\begin{array}
[c]{c}%
E_{t}x_{t+1}\\
E_{t}\pi_{t+1}%
\end{array}
\right)  =\underset{=\mathbf{A}_{yy}}{\underbrace{\left(
\begin{array}
[c]{cc}%
1+\frac{\gamma\kappa}{\beta} & -\frac{\gamma}{\beta}\\
-\frac{\kappa}{\beta} & \frac{1}{\beta}%
\end{array}
\right)  }}\underset{\mathbf{=y}_{t}}{\underbrace{\left(
\begin{array}
[c]{c}%
x_{t}\\
\pi_{t}%
\end{array}
\right)  }}+\underset{=\mathbf{B}_{y}}{\underbrace{\left(
\begin{array}
[c]{c}%
\gamma\\
0
\end{array}
\right)  }}i_{t}+\underset{=\mathbf{A}_{yz}}{\underbrace{\left(
\begin{array}
[c]{cc}%
-1 & \frac{\gamma}{\beta}\\
0 & -\frac{1}{\beta}%
\end{array}
\right)  }}\underset{\mathbf{=z}_{t}}{\underbrace{\left(
\begin{array}
[c]{c}%
z_{t}\\
u_{t}%
\end{array}
\right)  . }}%
\]

The matrix notation in bold below corresponds to the notation of the augmented
linear quadratic regulator (LQR) in Anderson, Hansen, McGrattan and Sargent
(1996, p.203) and Hansen and Sargent (2007). The two policy targets, the
output gap $x_{t}$ and inflation $\pi_{t}$, are two-time-step Kalman
controllable by a single policy instrument, the interest rate $i_{t}$ and its
first lag $i_{t-1}$, if \emph{both} the intertemporal elasticity of
substitution and the slope of the New-Keynesian Phillips curve are different
from zero: $\gamma\neq0$ which implies $\frac{\partial E_{t}\left(
x_{t+1}\right)  }{\partial i_{t}}\neq0$ and $\kappa=\frac{\partial
E_{t+1}\left(  \pi_{t+2}\right)  }{\partial x_{t+1}}\neq0$. This is found
computing the rank of Kalman controllability matrix:%

\begin{align*}
\text{rank}\left(  \mathbf{B}_{y},\mathbf{A}_{yy}\mathbf{B}_{y}\right)   &
=\text{rank}\left(  \left(
\begin{array}
[c]{c}%
\gamma\\
0
\end{array}
\right)  \text{, }\left(
\begin{array}
[c]{cc}%
1+\frac{\gamma\kappa}{\beta} & -\frac{\gamma}{\beta}\\
-\frac{\kappa}{\beta} & \frac{1}{\beta}%
\end{array}
\right)  \left(
\begin{array}
[c]{c}%
\gamma\\
0
\end{array}
\right)  \right) \\
&  =\text{rank}\left(
\begin{array}
[c]{cc}%
\gamma & \gamma\left(  \frac{\kappa}{\beta}\gamma+1\right) \\
0 & -\frac{\kappa}{\beta}\gamma
\end{array}
\right)  =2\text{ if }\gamma\neq0\text{ and }\kappa\neq0.
\end{align*}

If the intertemporal elasticity of substitution is zero $\gamma=0$, the rank
of the Kalman controllability matrix is zero. Then, the output gap is not
controllable by the interest rate in the first period. Even if the one-period
ahead output gap can control two-period ahead future inflation ($\kappa\neq
0$), as the current policy instrument (the interest rate) cannot control the
one-period ahead output gap, the interest rate cannot control the two-period
ahead inflation.

If the intertemporal elasticity of substitution is different from zero
$\gamma\neq0$ and $\kappa=0$, the rank of Kalman controllability matrix is
one. The output gap is controllable by the interest rate in the first period.
But the output gap next period cannot control two periods ahead inflation
($\kappa=0$), so that the current period interest rate cannot control two
periods ahead inflation.

If the intertemporal elasticity of substitution is different from zero
$\gamma\neq0$ and the slope of the New-Keynesian Phillips curve is different
from zero $\kappa\neq0$, the rank of Kalman controllability matrix is two. The
first period output gap is controllable by the interest rate. The first-period
ahead output gap controls two-period ahead inflation. Hence, there is a
non-zero correlation between expected two-period ahead inflation and current
interest, $\frac{\partial E_{t+1}\left(  \pi_{t+2}\right)  }{\partial i_{t}%
}=\frac{\partial E_{t+1}\left(  \pi_{t+2}\right)  }{\partial E_{t}\left(
x_{t+1}\right)  }\frac{\partial E_{t}\left(  x_{t+1}\right)  }{\partial i_{t}%
}\neq0$. The fact that the interest rate does not show up in the New-Keynesian
Phillips curve implies that the effect of the interest rate has to go through
a first-period effect on the output gap, so that a change on interest rate
only shows up in two-period ahead inflation. The positive correlation of the
interest rate on the one-time step expected output gap ($\gamma>0$) on the
first period is followed on the second-period negative correlation of the
one-period ahead output gap on two-period ahead expected inflation
($-\kappa>0$).%

\[
\frac{\partial E_{t+1}\left(  \pi_{t+2}\right)  }{\partial i_{t}%
}=\underset{<0}{\frac{\partial E_{t+1}\left(  \pi_{t+2}\right)  }{\partial
E_{t}\left(  x_{t+1}\right)  }}\underset{>0}{\frac{\partial E_{t}\left(
x_{t+1}\right)  }{\partial i_{t}}}<0.
\]

Kalman (1960) controllability is the generalization of Tinbergen's (1952)
principle for a static linear system of equations ($n$ policy instruments can
control $n$ policy targets in a \emph{single} period) to linear dynamic
systems. One policy instrument and its first lag can control two policy
targets in \emph{two} periods.

Secondly, the dynamics of the non-controllable exogenous forcing variables can
be stated as:%

\[
\left(
\begin{array}
[c]{c}%
z_{t+1}\\
u_{t+1}%
\end{array}
\right)  =\underset{=\mathbf{A}_{zz}}{\underbrace{\left(
\begin{array}
[c]{cc}%
\rho_{z} & 0\\
0 & \rho_{u}%
\end{array}
\right)  }}\left(
\begin{array}
[c]{c}%
z_{t}\\
u_{t}%
\end{array}
\right)  +\left(
\begin{array}
[c]{c}%
\varepsilon_{z,t}\\
\varepsilon_{u,t}%
\end{array}
\right)  .
\]

Thirdly, the feedback linear policy rule in the augmented linear quadratic
regulator responds also to forcing variables (Anderson, Hansen, McGrattan and
Sargent (1996, p.203), Hansen and Sargent (2007)):%

\[
i_{t}=\underset{=\mathbf{F}_{y}}{\underbrace{\left(
\begin{array}
[c]{cc}%
F_{x} & F_{\pi}%
\end{array}
\right)  }}\left(
\begin{array}
[c]{c}%
x_{t}\\
\pi_{t}%
\end{array}
\right)  +\underset{=\mathbf{F}_{z}}{\underbrace{\left(
\begin{array}
[c]{cc}%
F_{z} & F_{u}%
\end{array}
\right)  }}\left(
\begin{array}
[c]{c}%
z_{t}\\
u_{t}%
\end{array}
\right)  .
\]

\subsection{Dynamic Properties}

The dynamic properties of the New-Keynesian model depend on the eigenvalues of
the system linearized around the steady state. In this section we show that a
unique relationship between the two Taylor rule parameters $\left(  F_{\pi
},F_{x}\right)  $ and the trace and the determinant of the closed-loop matrix
exist. This will allow us to analyze determinacy and the existence of
bifurcations by looking at the Taylor rule parameters that have an economic
interpretation which is sometimes missing for the trace and determinant of a
dynamic system. Additionally to Barnett and Duzhak's (2008) Hopf bifurcation
border and Barnett and Duzhak's (2010) flip bifurcation border, the last
missing side of the stability triangle which corresponds to the saddle-node
bifurcation border can be identified. This specific relation in the
New-Keynesian model is a particular case of the mapping of feedback rule
parameters with closed-loop eigenvalues of controllable linear system
demonstrated by Wohnam (1967) pole placement theorem. We use Azariadis' (1993,
pp. 63-67) conditions on the trace and determinant for a discrete dynamic
system of dimension 2 in order to find the bifurcation limits of the stability
triangle in the plane of Taylor rule parameters.

The closed-loop matrix $\mathbf{A}_{yy}\mathbf{+B}_{y}\mathbf{F}_{y}$ (denoted
$\mathbf{A+BF}$ in what follows) of the controllable part of the New-Keynesian
model is:%

\[
\left(
\begin{array}
[c]{cc}%
1+\frac{\gamma\kappa}{\beta} & -\frac{\gamma}{\beta}\\
-\frac{\kappa}{\beta} & \frac{1}{\beta}%
\end{array}
\right)  +\left(
\begin{array}
[c]{c}%
\gamma\\
0
\end{array}
\right)  \left(
\begin{array}
[c]{cc}%
F_{x} & F_{\pi}%
\end{array}
\right)  =\left(
\begin{array}
[c]{cc}%
1+\frac{\gamma\kappa}{\beta}+\gamma F_{x} & -\frac{\gamma}{\beta}+\gamma
F_{\pi}\\
-\frac{\kappa}{\beta} & \frac{1}{\beta}%
\end{array}
\right)
\]

The characteristic polynomial of the closed-loop matrix $\mathbf{A+BF}$ of the
New-Keynesian model is a function of its trace $T$ and determinant $D$ leading
to two eigenvalues $\lambda_{1}$ and $\lambda_{2}$:%

\[
p(\lambda)=\det\left(  \mathbf{A+BF}-\lambda\mathbf{I}\right)  =\lambda
^{2}-T\lambda+D =(\lambda- \lambda_{1})(\lambda-\lambda_{2})=0.
\]

The eigenvalues (the roots of the characteristic polynomial) are non-linear
functions of the trace and the determinant. We either have two ordered real
eigenvalues or two complex conjugate eigenvalues:%

\begin{align*}
\lambda_{1}  &  =\frac{T-\sqrt{T^{2}-4D}}{2}<\text{ }\lambda_{2}=\frac
{T+\sqrt{T^{2}-4D}}{2}\text{ \ \ if }\Delta=T^{2}-4D\geq0\\
\text{ }\lambda_{1}  &  =\frac{T-i\sqrt{4D-T^{2}}}{2}\text{ and }\lambda
_{2}=\overline{\lambda}_{1}=\frac{T+i\sqrt{4D-T^{2}}}{2}\text{ \ if }%
\Delta=T^{2}-4D<0.
\end{align*}

The trace $T$ and determinant $D$ of the closed-loop matrix $\mathbf{A+BF}$,
however, are affine functions of the Taylor rule parameters $F_{x}$ and
$F_{\pi}$, where constants are, respectively, the trace $T\left(
\mathbf{A}\right)  $ and the determinant $D\left(  \mathbf{A}\right)  $ of the
open-loop matrix $\mathbf{A}$:%

\begin{align}
T\left(  \mathbf{A+BF}\right)   &  =1+\frac{1}{\beta}+\frac{\gamma\kappa
}{\beta}+\gamma F_{x}=T\left(  \mathbf{A}\right)  +\gamma F_{x}\text{ ,}\\
D\left(  \mathbf{A+BF}\right)   &  =\frac{1}{\beta}+\frac{\gamma}{\beta}%
F_{x}+\frac{\gamma\kappa}{\beta}F_{\pi}=D\left(  \mathbf{A}\right)  +\frac
{1}{\beta}\gamma F_{x}+\frac{\kappa}{\beta}\gamma F_{\pi},\\
\text{with }T\left(  \mathbf{A}\right)   &  =1+\frac{1}{\beta}+\frac
{\gamma\kappa}{\beta}\text{ and }D\left(  \mathbf{A}\right)  =\frac{1}{\beta}.
\end{align}

Conversely, the Taylor rule parameters $F_{x}$ and $F_{\pi}$ are linear
functions of the difference of the closed-loop and the open-loop trace
$T\left(  \mathbf{A+BF}\right)  -T\left(  \mathbf{A}\right)  $ and of the
difference of the closed-loop and the open-loop determinant $D\left(
\mathbf{A+BF}\right)  -D\left(  \mathbf{A}\right)  $ and affine functions of
the closed-loop trace and determinant:%

\begin{align}
F_{x}  &  =\frac{1}{\gamma}\left(  T\left(  \mathbf{A+BF}\right)  -T\left(
\mathbf{A}\right)  \right)  = \frac{1}{\gamma}T\left(  \mathbf{A+BF}\right)
-\frac{1}{\gamma}\left(  1+\frac{1}{\beta}+\frac{\gamma\kappa}{\beta}\right)
,\\
F_{\pi}  &  =-\frac{1}{\gamma\kappa}\left(  T\left(  \mathbf{A+BF}\right)
-T\left(  \mathbf{A}\right)  \right)  +\frac{\beta}{\gamma\kappa}\left(
D\left(  \mathbf{A+BF}\right)  -D\left(  \mathbf{A}\right)  \right) \\
&  =-\frac{1}{\gamma\kappa}T\left(  \mathbf{A+BF}\right)  +\frac{\beta}%
{\gamma\kappa}D\left(  \mathbf{A+BF}\right)  +\frac{1}{\beta}+\frac{1}%
{\gamma\kappa\beta} \text{ .}\nonumber
\end{align}

In classic control, these feedback rule parameters are corresponding to a
\textquotedblleft pole placement" with desired closed-loop eigenvalues defined
by the parameters of the characteristic polynomial, which are in dimension 2,
the trace $T\left(  \mathbf{A+BF}\right)  $ and the determinant $D\left(
\mathbf{A+BF}\right)  $. This relation can also be found using the canonical
form of the dynamic system or using Ackermann's (1972) formula (see appendix B).

In Azariadis (1993, chapter 8), a stability triangle with Hopf ($D=1$ and
$\Delta<0$), flip ($p(-1)=0$) and saddle-node ($p(1)=0$) bifurcation borders
and a quadratic function delimiting complex conjugate versus non-complex
solutions (discriminant $\Delta=0$) are described in the plane of the trace
and determinant $\left(  T,D\right)  $. This defines areas with zero, one or
two stable eigenvalues $\left(  \lambda_{1},\lambda_{2}\right)  $ in the plane
$\left(  T,D\right)  $. We use Azariadis' (1993) insights to compute
bifurcation borders in the plane of Taylor rule bifurcation parameters
$\left(  F_{\pi},F_{x}\right)  $. An example for given parameter values of the
New-Keynesian model is depicted in Figure 1 with the numerical values for the
corresponding points and the values of the rule parameters and the eigenvalues
given in Table 1. Inside the triangle ABC the eigenvalues have modulus smaller
than 1, and outside the triangle at least one eigenvalue is larger than one in
absolute value. The dotted parabola through the points A, $\Omega$, and B
corresponds to a discriminant equal to zero with complex eigenvalues on the
right-hand side of the parabola and real eigenvalues on its left-hand side.
The center $\Omega$ has both eigenvalues equal to zero ($\lambda_{1}%
=\lambda_{2}=0$) and provides the Taylor rule parameters with the fastest
stabilization.
\[
\Delta=0\Leftrightarrow\frac{1}{\beta}+\frac{\gamma}{\beta}F_{x}+\frac
{\gamma\kappa}{\beta}F_{\pi}=\frac{1}{4}\left(  1+\frac{1}{\beta}+\frac
{\gamma\kappa}{\beta}+\gamma F_{x}\right)  ^{2}.
\]
On the segment connecting the points A and B both eigenvalues are conjugate
complex and have modulus one. This is the case when the determinant is equal
to one:
\[
D(\mathbf{A+BF})=1\Leftrightarrow F_{x}=\frac{\beta-1}{\gamma}-\kappa F_{\pi}.
\]
Above this line, that is for policy parameters $F_{x}>\frac{\beta-1}{\gamma
}-\kappa F_{\pi}$, the eigenvalues have modulus larger than one and below this
line, that is for policy parameters $F_{x}<\frac{\beta-1}{\gamma}-\kappa
F_{\pi}$, the eigenvalues have modulus smaller than one. Crossing this line
from inside the triangle corresponds to a Hopf bifurcation, see also Barnett
and Duzhak (2008).

To analyze the real eigenvalues, we look at the characteristic polynomial:
\begin{align}
p(a) &  =\left(  a-\lambda_{1}\right)  \left(  a-\lambda_{2}\right)
>0\Leftrightarrow a<\lambda_{1}<\lambda_{2}\text{ or }\lambda_{1}<\lambda
_{2}<a,\\
\text{ }p(a) &  =\left(  a-\lambda_{1}\right)  \left(  a-\lambda_{2}\right)
<0\Leftrightarrow\lambda_{1}<a<\lambda_{2}\text{ .}%
\end{align}
The two lines $p\left(  1\right)  =0$ and $p\left(  -1\right)  =0$ divide the
plane into four regions as shown in Figures 1 and 2. They cross at point C
($\lambda_{1}=-1$ and $\lambda_{2}=1$).

On the line going through the points A and C, at least one of the real
eigenvalues is equal to one. It is characterized by
\begin{align*}
p\left(  1\right)   &  =1-T(\mathbf{A+BF})+D(\mathbf{A+BF})=0\\
\Leftrightarrow F_{x} &  =\frac{\kappa}{1-\beta}\left(  1-F_{\pi}\right)
\text{ or }F_{\pi}=1-\left(  \frac{1-\beta}{\kappa}\right)  F_{x}\text{. }%
\end{align*}

If $F_{x}>0$ but not too large, $F_{\pi}$ can be slightly below one. If
$F_{x}<0$, $F_{\pi}$ is strictly larger than one: it satisfies the Taylor
principle. When the discount factor $\beta$ is close to one, the slope of the
line is close to infinity (vertical line), so that this condition is close to
the Taylor principle condition $F_{\pi}>1$. It goes through the point A
($\lambda_{1}=\lambda_{2}=1$) of Table 1\ which is the point of intersection
of the parabola of a discriminant equal to zero ($\Delta=0$) and a determinant
equal to one $D=\lambda_{1}\lambda_{2}=1$. It goes through the point C
($\lambda_{1}=-1$ and $\lambda_{2}=1$) of Table 1, point of intersection of
the line $p\left(  -1\right)  =0$.

On the right-hand side of the line $p\left(  1\right)  =0$ and inside the
triangle, both eigenvalues are smaller than one in absolute value. Crossing
the line from the left leads to one eigenvalue larger than one and the other
one smaller than one, a saddle-node bifurcation.

Similarly, \ %

\begin{equation}
p\left(  -1\right)  =1+T+D=0\Rightarrow F_{x}=-\frac{2}{\gamma}-\frac{\kappa
}{1+\beta}-\frac{\kappa}{1+\beta}F_{\pi}\text{ }%
\end{equation}

which goes through the points C ($\lambda_{1}=-1$ and $\lambda_{2}=1$) and B
($\lambda_{1}=\lambda_{2}=-1$), see Table 1. Above this line and inside the
triangle, we have $p\left(  -1\right)  >0$, with both roots smaller than $1$
in absolute value. Below this line $p\left(  -1\right)  <0$, one of the
eigenvalues is smaller than $-1$ and the other eigenvalue is larger than $-1$.
Crossing this line from inside the triangle corresponds to a flip bifurcation,
see also Barnett and Duzhak (2010).

\ 

The parameter values in Table 1 correspond to some estimated values:
$\gamma=0.5$ for the intertemporal elasticity in the USA (Havranek \textit{et
al.} (2015)) and $\kappa=0.1$ for the slope of the New-Keynesian Phillips
curve $\kappa=0.1$ (Mavroeidis \textit{et al.} (2014)).

\textbf{Table 1: Stability triangle with center and point of laissez-faire
(origin) of the New-Keynesian model (}$\gamma=0.5$\textbf{, }$\kappa
=0.1$,$\beta=0.99$\textbf{), with }$T\left(  \mathbf{A}\right)  =1+\frac
{1}{\beta}+\frac{\gamma\kappa}{\beta}=2.\,\allowbreak06>2$.%

\begin{tabular}
[c]{|l|l|l|l|l|l|l|}\hline
& $\lambda_{1}$ & $\lambda_{2}$ & $\lambda_{1}+\lambda_{2}$ & $\lambda
_{1}\lambda_{2}$ & $F_{\pi}$ & $F_{x}$\\\hline
A & $1$ & $1$ & $2$ & $1$ & $\frac{-\left(  2-T\left(  \mathbf{A}\right)
\right)  +\beta\left(  1-\frac{1}{\beta}\right)  }{\gamma\kappa}=1.01$ &
$\frac{2-T\left(  \mathbf{A}\right)  }{\gamma}=-0.12$\\\hline
B & $-1$ & $-1$ & $-2$ & $1$ & $\frac{-\left(  -2-T\left(  \mathbf{A}\right)
\right)  +\beta\left(  1-\frac{1}{\beta}\right)  }{\gamma\kappa}=81.0$ &
$\frac{-2-T\left(  \mathbf{A}\right)  }{\gamma}=-8.12$\\\hline
C & $-1$ & $1$ & $0$ & $-1$ & $\frac{T\left(  \mathbf{A}\right)  +\beta\left(
-1-\frac{1}{\beta}\right)  }{\gamma\kappa}=1.41$ & $-\frac{T\left(
\mathbf{A}\right)  }{\gamma}=-4.12$\\\hline
$\Omega$ & $0$ & $0$ & $0$ & $0$ & $\frac{T\left(  \mathbf{A}\right)
+\beta\left(  -\frac{1}{\beta}\right)  }{\gamma\kappa}=21$ & $-\frac{T\left(
\mathbf{A}\right)  }{\gamma}=-4.12$\\\hline
O & $0<\lambda_{1}<1$ & $1<\frac{1}{\lambda_{1}\beta}$ & $T\left(
\mathbf{A}\right)  $ & $\frac{1}{\beta}$ & $0$ & $\frac{T\left(
\mathbf{A}\right)  -T\left(  \mathbf{A}\right)  }{\gamma}=0$\\\hline
\end{tabular}

Figure 1 represents bifurcation lines delimiting the number of stable
eigenvalues for the New-Keynesian model for positive sign restrictions on
monetary policy transmission mechanism. The case of negative sign restrictions
on monetary policy transmission mechanism is treated separately in section 2.4
(Figure 2). Figure 1\ shows how the plane is divided by the four reference
lines $p\left(  -1\right)  =0$, $p\left(  1\right)  =0$, $\Delta=0$ and $D=1$
into eight regions. We start in the upper left quadrant and move counter-clockwise.

\textit{Region 1:} On the left of the line AC and above the line BC, $p\left(
1\right)  <0$ and $p(-1)>0$. Both eigenvalues are on the same side of $-1$ and
on different sides of $1$. The only possibility is $-1<\lambda_{1}%
<1<\lambda_{2}$. The steady state is a saddle-point. This area includes the
origin $0$ which corresponds to the laissez-faire open-loop equilibrium, where
both Taylor rule parameters are equal to zero.

\textit{Region 2:} On the left of the line AC and below the line BC, $p\left(
1\right)  <0$ and $p(-1)<0$. The only possibility is $\lambda_{1}<-1$ and
$1<\lambda_{2}$. Both eigenvalues are on different sides of $1$ and on
different sides of $-1$. The steady state is a source.

\textit{Region 3:} On the right of the line AC and below the line BC,
$p\left(  1\right)  >0$ and $p(-1)<0$. Both eigenvalues are on the same side
of $1$ and on different sides of $-1$. The only possibility is $\lambda
_{1}<-1<\lambda_{2}<1$. The steady state is a saddle-point.

\textit{Region 4:} On the right of the line AC and above the line BC,
$p\left(  1\right)  >0$ and $p(-1)>0$. This region is split into five regions
as follows:

(4.1): Above the line BC and below the parabola: Both eigenvalues are real
with $\ -1<\lambda_{1}<\lambda_{2}<1$. The steady state is a sink.

(4.2): Above the parabola and below the line AB: Both eigenvalues are
conjugate complex with modulus smaller than one $\left\vert \lambda
_{1}\right\vert =\left\vert \lambda_{2}\right\vert <1$, the steady state is a sink.

(4.3): Above the line AB and inside the parabola: Both eigenvalues are
conjugate complex with modulus larger than one $\left\vert \lambda
_{1}\right\vert =\left\vert \lambda_{2}\right\vert >1$, the steady state is a source.

(4.4): Outside the parabola and on the right of the line AC: Both eigenvalues
are real and larger than one with $1<\lambda_{1}<\lambda_{2}$. The steady
state is a source.

(4.5): The last region corresponds to the area outside the triangle above the
line BC below the parabola. Here, both eigenvalues are negative $\lambda
_{1}<\lambda_{2}<-1$, with their sum (trace) below $-2$.

In the stability triangle ABC, the Taylor principle is satisfied $F_{\pi}>1$
and the output gap Taylor rule parameter is negative: $F_{x}<0$.

Crossing the lines $p(1)=0$ or $p(-1)=0$ outside the stability triangle
\emph{corresponds to another type of bifurcations between unstable dynamic
systems} in the sense that number of stable eigenvalues shifts from one to
zero or the reverse. In particular, for positive Taylor rule parameters
$F_{\pi}\geq0$ and $F_{x}\geq0$, the laissez-faire regime with zero Taylor
rule parameters (the origin of Figure 1) corresponds to a case such that
$-1<\lambda_{1}<1<\lambda_{2}$, to the left of the line $p(1)=0$. This
laissez-faire ("open-loop") private sector's model is described by the Fed
following a fixed interest rate target or peg: $i_{t}-i^{\ast}=0$. An increase
of the inflation Taylor rule parameter $F_{\pi}$ shifts the dynamic system
from the area where $-1<\lambda_{1}<1<\lambda_{2}$ with $p(1)<0$ up to
crossing the line $p(1)=0$ to a an area where $1<\lambda_{1}<\lambda_{2}$ with
$p(1)>0$. This is a saddle-node bifurcation between unstable systems which are
both having at least one unstable eigenvalue $\lambda_{2}$.

\textbf{Proposition 1. }\textit{For strictly positive intertemporal elasticity
of substitution }$\gamma>0$\textit{ and a strictly positive slope of the
New-Keynesian Phillips curve }$\kappa>0$\textit{ and a positive discount
factor below one }$0<\beta\leq1$\textit{, a necessary condition for having two
stable eigenvalues (}$\left\vert \lambda_{1}\right\vert <1$\textit{ and
}$\left\vert \lambda_{2}\right\vert <1$\textit{) in the closed-loop transition
matrix of the New-Keynesian model is that rule parameters }$\left(
F_{x},F_{\pi}\right)  $\textit{ are inside in a stability triangle ABC such
that the output gap parameter is strictly negative }$\left(  F_{x}<0\right)
$\textit{ and the inflation rule parameter is larger than one }$\left(
F_{\pi}>1-\left(  \frac{1-\beta}{\kappa}\right)  F_{x}>1\text{ or
}p(1)>0\right)  $\textit{, according to the Taylor principle.}

\textbf{Proof}.\textbf{ }Eigenvalues are both stable ($\left\vert \lambda
_{1}\right\vert <1$ and $\left\vert \lambda_{2}\right\vert <1$) if and only if
$p(-1)>0$, $p(1)>0$ and $D<1$. The condition $p(1)>0$ provides the Taylor
principle condition. The negative condition on the output gap Taylor rule
parameter is obtained as follows. The maximal value of the output gap Taylor
rule parameter is necessarily on one of the apexes of the stability triangle
ABC which corresponds to intersections of $p(-1)>0$, $p(1)>0$ and $D<1$. It
turns out that the values of this parameter for the three points defining the
triangle are ordered as follows and that the largest one $F_{x}(A)$ is
strictly negative:%
\[
F_{x}(B)<F_{x}(C)<F_{x}(A)<0.
\]

Because $\gamma>0$, $\kappa>0$ and $0<\beta\leq1$, one has:%
\[
\frac{-2-T\left(  \mathbf{A}\right)  }{\gamma}<-\frac{T\left(  \mathbf{A}%
\right)  }{\gamma}<\frac{2-T\left(  \mathbf{A}\right)  }{\gamma}<0.
\]

In particular, the largest value of the output gap Taylor rule parameter is
negative $F_{x}(A)<0$ because $\gamma\kappa>0$ and $0<\beta\leq1$:%

\[
\text{If }\gamma>0\text{, }\kappa>0\text{ and }0<\beta\leq1\text{ then }%
F_{x}(A)<0\Leftrightarrow2<T\left(  \mathbf{A}\right)  =1+\frac{1+\gamma
\kappa}{\beta}.
\]

Q.E.D.

\textbf{Proposition 2. }If $\gamma>0$, $\kappa>0$ and $0<\beta\leq1$\textit{,
then for policy rule parameters within the stability triangle ABC
}($\left\vert \lambda_{1}\right\vert <1$ \textit{and} $\left\vert \lambda
_{2}\right\vert <1$ \textit{if and only if} $p(-1)>0$, $p(1)>0$ and
$D<1$)\textit{, the trace and determinant of the closed-loop system are lower
than in the case of the open-loop laissez-faire system, where }$0<\lambda
_{1}<1<\frac{1}{\lambda_{1}\beta}=\lambda_{2}$:\textit{ }$T\left(
\mathbf{A+BF}\right)  <T\left(  \mathbf{A}\right)  $ and $D\left(
\mathbf{A+BF}\right)  <1\leq D\left(  \mathbf{A}\right)  =1/\beta$.

\textbf{Proof: }As $F_{x}=\frac{T\left(  \mathbf{A+BF}\right)  -T\left(
\mathbf{A}\right)  }{\gamma}<0$ in the stability triangle ABC, this implies
that $T\left(  \mathbf{A+BF}\right)  <T\left(  \mathbf{A}\right)  $. A
condition of the stability triangle is that the product of eigenvalues is
smaller than one: $D\left(  \mathbf{A+BF}\right)  <1$. By contrast, the
open-loop system determinant is the inverse of the discount factor, which is
at least equal to one: $1\leq D\left(  \mathbf{A}\right)  =1/\beta$, hence
$D\left(  \mathbf{A+BF}\right)  <1\leq D\left(  \mathbf{A}\right)  =1/\beta$.

Q.E.D.

\subsection{Negative versus Positive Feedback}

The fact that a negative output gap Taylor rule parameter is a necessary
condition for the closed-loop dynamic system to have two eigenvalues within
the unit circle, requires further investigation. In this section, we explain
how to distinguish negative feedback from positive feedback in the scalar
case. We explain how the sign of the negative feedback policy rule parameter
changes when the sign of the transmission mechanism changes. We are then able
to explain that, because of the intertemporal substitution effect of the
interest rate on consumption in the transmission mechanism, a negative output
gap Taylor rule parameter corresponds to a negative feedback mechanism in the
Taylor rule. By contrast, with a delayed cost-of-capital mechanism, a positive
output gap Taylor rule parameter corresponds to a negative feedback mechanism
of the Taylor rule.

Let us consider any first order scalar case transmission mechanism model and a
feedback rule for a policy target $x_{t}$ and a policy instrument $i_{t}$
(first order "single input, single output", "SISO"\ case) with $A$, $B$ and
$F$ real numbers:%
\begin{align*}
x_{t+1} &  =Ax_{t}+Bi_{t}\text{: Transmission mechanism; }i_{t}=Fx_{t}\text{:
Feedback rule}\\
x_{t+1} &  =(A+BF)x_{t}\text{: Closed loop system; }x_{t+1}=Ax_{t}\text{:
open-loop system (}F=0\text{).}%
\end{align*}

The closed-loop system is obtained after substitution of the feedback rule.
The open-loop system (or "laissez-faire") corresponds to no policy
intervention: the feedback rule parameter $F$ is equal to zero.

\textbf{Definition 3:} \textit{A negative feedback stabilization mechanism is
defined as a set of stabilizing policy parameters such that the expectation of
the gap of the target with respect to its long-run equilibrium value with
policy intervention is smaller than the expectation of the gap of the target
without intervention.}

In the scalar case with a single input and a single output, the negative
feedback stabilization mechanism is obtained by the following set of policy
rule parameters $F$:%

\[
\left\{  F\in%
\mathbb{R}
,\text{ with }F\text{ such that }\left\vert A+BF\right\vert <\min\left(
1,\left\vert A\right\vert \right)  \right\}  .
\]

A negative feedback mechanism implies that $\left\vert A+BF\right\vert
<\left\vert A\right\vert $, the policy rule parameter decreases the eigenvalue
of the closed-loop dynamic system with respect to the open-loop dynamic
system. It accelerates convergence and decreases the endogenous persistence
(auto-correlation $\left\vert A+BF\right\vert $) of the policy target $x_{t}$
with respect to its open-loop persistence measured by $\left\vert A\right\vert
$. However, this may not lead to stabilization in the case where both
$\left\vert A+BF\right\vert $ and $\left\vert A\right\vert $ are larger than
one. In order to have negative feedback \textit{and} stabilization, one needs
the closed-loop eigenvalue to be below one in absolute value: $\left\vert
A+BF\right\vert <1<\left\vert A\right\vert $. In this case, moving from
negative feedback with stabilization to \textquotedblleft
laissez-faire\textquotedblright\ corresponds to a saddle-node bifurcation if
$A$ is larger than 1, $-1<A+BF<1<A$, and to a a flip bifurcation if $A$ is
lower than $-1$, $A<-1<A+BF<1$. In the case $\left\vert A+BF\right\vert
<\left\vert A\right\vert <1$, we have a negative feedback mechanism, but there
is no bifurcation because the open-loop system is already stable (its
eigenvalue $\left\vert A\right\vert $ is in the unit circle).

\textbf{Proposition 4.} \emph{In the scalar case, for a positive open-loop
auto-correlation of the policy target }$A>0$\emph{, a change of sign of the
marginal correlation }$B$\emph{ of the current policy instrument }$i_{t}%
$\emph{ on the future value of the policy target }$\emph{x}_{t+1}$\emph{ in
the monetary policy transmission mechanism implies a change of sign of the
negative feedback rule parameter }$F$\emph{: }$B>0$\emph{ implies }$F<0$\emph{
or }$B<0$\emph{ implies }$F>0$\emph{, because negative feedback implies that
}$BF<0$.

\textbf{Proof.}%

\begin{align}
\left\vert A+BF\right\vert  &  <A\Rightarrow-A<A+BF<A\Rightarrow-2A<BF<0.\\
\text{If }B  &  >0\Rightarrow-2\frac{A}{B}<F<0\text{ else if }B<0\Rightarrow
0<F<-2\frac{A}{B}\text{ .}\nonumber
\end{align}

QED.

In other words, inspecting only of the sign of the parameter $F$ of the
proportional feedback rule is not a proof of negative feedback mechanism, one
needs to inspect the sign of the transmission mechanism parameter $B=$ and
check that $BF<0$.

negative feedback implies \emph{bounded }feedback rule parameters on\emph{
both }sides, and not only an asymmetric boundary condition such as the Taylor
principle ($F_{\pi}>1$). An excessive opposite feedback reaction would lead to
a lower closed-loop eigenvalue than an opposite sign of the open-loop
eigenvalue $A+BF<-A$ if $A>0$. It could lead to instability in the case where
$A+BF<-1<-A$, if $A>0$.

When a loss function includes a quadratic cost on the volatility of the policy
instrument, the reduced form of the optimal policy feedback rule parameter $F$
does not change the sign of the eigenvalue with respect to the open-loop
eigenvalue. Else, it would lead to more volatility of the policy instrument
for the same absolute value of the eigenvalue of the policy target. Hence, the
range of variation of the policy instrument is such that $0\leq A+BF<A$ and
$-2A<-A\leq BF<0$ (see Chatelain and Ralf (2016) for Ramsey optimal policy
with a scalar case model). In this paper, the area of Ramsey optimal policy
reduced form rule parameters remains on the same side of the center of the
stability triangle (see Ramsey optimal policy section).

\textbf{Proposition 5. }If $\gamma>0$, $\kappa>0$ and $0<\beta\leq1$\textit{,
}\emph{a negative Taylor rule parameter on the output gap }$F_{x}<0$\emph{ is
a necessary condition for a negative feedback mechanism in the New-Keynesian
model.}

\textbf{Proof. }The New-Keynesian transition matrix is:%

\[
\left(
\begin{array}
[c]{c}%
E_{t}x_{t+1}\\
E_{t}\pi_{t+1}%
\end{array}
\right)  =\left(
\begin{array}
[c]{cc}%
1+\frac{\gamma\kappa}{\beta}+\gamma F_{x} & -\frac{\gamma}{\beta}+\gamma
F_{\pi}\\
-\frac{\kappa}{\beta} & \frac{1}{\beta}%
\end{array}
\right)  \left(
\begin{array}
[c]{c}%
x_{t}\\
\pi_{t}%
\end{array}
\right)  .
\]

Let us consider a particular initial output gap shock where $x_{0}>0$ and
$\pi_{0}=0$. We consider this particular case, because we only seek a
necessary condition in order to provide an economic intuition explaining why
the output gap rule parameter is negative for negative feedback. The necessary
and sufficient conditions have been given in proposition 1: they define the
stability triangle.

With the output gap shock where $x_{0}>0$ and $\pi_{0}=0$, the output equation
corresponds to the scalar case. Negative feedback mechanism implies that the
expectation of the output gap with policy intervention $E_{0}x_{1}$ should be
lower than the expectation of next period's output gap without policy
intervention $E_{0}x_{1,F_{x}=0}$:%
\[
E_{0}x_{1}=\left(  1+\frac{\gamma\kappa}{\beta}+\gamma F_{x}\right)
x_{0}<E_{0}x_{1,F_{x}=0}=\left(  1+\frac{\gamma\kappa}{\beta}\right)
x_{0}\Leftrightarrow\gamma F_{x}<0
\]

Since $\gamma>0$ this implies $F_{x}<0$. \emph{Because of the intertemporal
substitution effect on consumption} $B=\gamma>0$, \emph{a rise of the interest
rate goes hand in hand with a rise of future output gap.} Hence, negative
feedback implies a negative output gap-Taylor-rule parameter $F=F_{x}<0$.

Q.E.D.

\subsection{Alternative Monetary Policy Transmission Mechanism}

This section seeks an alternative ad hoc monetary policy transmission
mechanism such that negative feedback is compatible with a positive output gap
rule parameter and a positive inflation Taylor rule parameter.

In the New-Keynesian model, the production function does not depend on
capital. Hence, the monetary policy transmission channel of the cost of
capital is assumed away. Central bankers often believe that, for a positive
output gap, a rise of the interest rate leads to a fall of current output gap
which leads to a fall of future inflation. For example, Taylor (1999) assumes
no intertemporal substitution effect $(\gamma=0)$ but a negative effect of the
\emph{user cost of capital }on current output taking into account current
inflation instead of expected inflation in the cost of capital. He also
assumes an accelerationist Phillips curve effect, with an opposite negative
correlation of future inflation with current output:
\[
x_{t}=-a\left(  i_{t}-\pi_{t}\right)  \text{ and }\pi_{t+1}=\pi_{t}+bx_{t}%
=\pi_{t}-ab\left(  i_{t}-\pi_{t}\right)  \text{, with }a>0\text{, }%
b=-\kappa>0.
\]

This dynamic system has only one dimension. It is similar to the extreme case
of the New-Keynesian Phillips curve where all labor is financed by working
capital so that $\kappa_{i}=ab>0$ (Christiano \textit{et al.} (2010),
Bratsiotis and Robinson (2016)). Future inflation is negatively correlated
with the cost of working capital and not at all with the cost of labor.
Because there is only one recursive dimension, the Taylor rule responds only
to inflation: $i_{t}=F_{\pi}\pi_{t}$. The closed-loop system converges faster
than the open-loop (negative feedback) if and only if the inflation Taylor
rule parameter is larger than one (Taylor principle) and bounded from above:%

\begin{align*}
-\pi_{t}  &  <\pi_{t+1}=\left(  1+ab-abF_{\pi}\right)  \pi_{t}<\pi
_{t}\Leftrightarrow-2-ab<-abF_{\pi}<-ab\Leftrightarrow\\
1  &  <F_{\pi}<1+\frac{2}{ab}.
\end{align*}

Now, let us introduce a second recursive dimension for the output gap besides
inflation dynamics, namely an \emph{ad hoc} delayed cost-of-capital effect. An
increase of the interest rate leads to a decrease of \emph{future} production
$(\gamma<0)$. Based on microeconomic foundations assuming that some agents
face credit rationing or limited asset market participation, this negative
aggregate intertemporal elasticity of substitution is found in heterogeneous
agents New-Keynesian models (Bilbiie (2008), Bilbiie and Straub (2013)).
Empirically, Havranek (2015) and Havranek \textit{et al.} (2015) found near
zero estimates at macro level or negative estimates of $\gamma$ for some
countries in their meta-analysis. Mavroeidis \textit{et al.} (2014) found
hundreds of negative estimates of $\kappa$ using USA data.

\textbf{Proposition 6. }\textit{For }$0<\beta\leq1$\textit{ and for the
alternative transmission mechanism, }$\gamma<0$\textit{, }$\kappa<0$\textit{,
a necessary condition for having two stable eigenvalues (}$\left\vert
\lambda_{1}\right\vert <1$\textit{ and }$\left\vert \lambda_{2}\right\vert
<1$\textit{) in the closed-loop transition matrix of the New-Keynesian model
is that rule parameters }$\left(  F_{x},F_{\pi}\right)  $\textit{ are inside a
stability triangle such that the output gap parameter is strictly
\emph{positive} }$\left(  F_{x}>0\right)  $\textit{ and the inflation rule
parameter is above one }$\left(  F_{\pi}>1-\left(  \frac{1-\beta}{\kappa
}\right)  F_{x}\right)  $\textit{, according to the Taylor principle. }

\textbf{Proof. }Same computations as for Proposition 2 with opposite signs for
$\gamma$ and $\kappa$.

The stability triangle in Figure 2 with opposite sign restrictions $\gamma<0$
and $\kappa<0$, is symmetric to the stability triangle of Figure 1 with
respect to the horizontal axis in the positive quadrant of ($F_{x}$,$F_{\pi}%
$). The economic intuition is that a rise of the interest rate today implies a
fall of next period's output which implies a fall of inflation two-period from
now. Then, negative feedback implies a positive output gap Taylor rule parameter.

\subsection{Determinacy with Forward-looking Interest Rate}

This section investigates why determinacy in the New-Keynesian model with a
Taylor rule seeks two unstable eigenvalues and hence recommends a positive
feedback mechanism for the central bank to stabilize the output gap and inflation.

Blanchard and Kahn's (1980) condition for the solution of a linear rational
expectations dynamic system to be determinate (unique) is that the number of
predetermined variables should be equal to the number of stable eigenvalues.
In this paper, which variables are backward-looking (predetermined) and which
are forward-looking, however, is an \emph{ad hoc choice of the researcher} as
it was the rule in \emph{ad hoc} rational expectations models of the 1970s
(Chatelain and Ralf (2019c)).

Blanchard and Kahn (1980) acknowledge that their determinacy condition is
always satisfied for optimal growth and optimal control with a linear
quadratic regulator when concavity conditions are satisfied. In this case, the
\emph{ad hoc} matrix of the linear rational expectations system is replaced by
the symplectic matrix of a Hamiltonian system \emph{derived from optimal
conditions} with Lagrange multipliers. A symplectic matrix (its transpose is
similar to its inverse) is such that the number of unstable eigenvalues is
equal to the number of stable eigenvalues. The fact that a state variable or a
costate variable is forward-looking or predetermined is \emph{endogenously
derived} from optimal control or the dynamic Stackelberg game program and its
transversality conditions. The number of predetermined variables is equal to
the number of forward-looking variables.

The New-Keynesian model is a hybrid model of optimal behavior for the private
sector and of \emph{ad hoc} behavior for the policy maker described by a
policy instrument (the interest rate) which reacts to private sector variables
according to an ad hoc feedback Taylor rule. The fact that the output gap and
inflation are forward-looking is endogenously derived from the private
sector's optimal control program. By contrast, the assumption that the
interest rate and its lag are forward-looking variables is an ad hoc choice by
a researcher. A researcher could as well select the opposite assumption that
the interest and its lag are predetermined variables. It is only an arbitrary
assumption as it was done in \emph{ad hoc} rational expectations models in the 1970s.

Using only Blanchard and Kahn's (1980) condition, it is not possible to prove
that policy maker's instruments of feedback rules are necessarily
forward-looking. To prove this would require a model of intertemporal optimal
behavior by the policy maker. But, consistent with negative feedback
mechanism, Ramsey optimal policy proves the opposite: policy instruments are
backward-looking when policy targets are forward-looking.\emph{ }

In the New-Keynesian model, the two policy targets (inflation and output gap)
are forward-looking variables, \emph{because} they are costate variables of
private sector optimal intertemporal behavior. In a second step, the
New-Keynesian theory uses the \emph{ad hoc} assumption that the policy
instrument (interest rate) in the Taylor rule is a forward-looking variable.
As seen in the next section, Ramsey optimal policy implies something else: the
policy instrument (interest rate) and its lag are backward-looking variables,
\emph{because} the two policy targets (inflation and output gap) are
forward-looking variables.

Once it is assumed that the interest rate is forward-looking as well as
inflation and the output gap, and once it is assumed that the predetermined
variable of the private sector (wealth or public debt) is set to zero in all
periods (Gali (2015), footnote 3), the model is a degenerate rational
expectations model without predetermined variable. In this case, any shock
leads to an instantaneous jump back to the long-run equilibrium value of
inflation and output which is the only available anchor. Hence, there is no
transitory dynamics. There is neither a need for stabilization policy nor for
a feedback Taylor rule.

In order to introduce transitory dynamics, a New-Keynesian \emph{convention}
is to add \emph{ad hoc} autoregressive components to the cost-push shock and
to the output gap shock. The expectations of these two autoregressive
components provide two predetermined variables with transitory dynamics on
which inflation and the output gap can be anchored. The fact that these
anchors are not observable may be worrisome for central bank practitioners who
usually anchor inflation on observable variables, such as their policy instruments.

Once the decision is taken by the researcher on which variables are
forward-looking, Blanchard and Kahn's (1980) determinacy condition for ad hoc
rational expectations linear systems is applied mechanically. The number of
stable eigenvalues should be equal to the number of predetermined variables.
There are two exogenous predetermined variables with their own stable
eigenvalues corresponding to their auto-correlation parameters. There are
three endogenous forward-looking variables (output gap, inflation and the
interest rate) whose dynamics corresponds to a closed-loop endogenous system
including a feedback Taylor rule in two dimensions. This implies that this two
dimensional system should have two unstable eigenvalues. Hence, the
determinacy area in the plane of the policy rule parameters corresponds to the
areas where there are two unstable eigenvalues (regions 4.3 and 4.4 with $D>1$
and region 2 in Figure 1).

Because the Taylor rule parameters determine the stability or instability of
eigenvalues, as a particular case of Wonham's (1967) pole placement theorem,
when they force eigenvalues to be unstable instead of stable, they correspond
to a positive feedback mechanism in the sense of control.

The determinacy area includes \emph{unlikely} unbounded areas which are
allowed by the positive feedback mechanism following the arbitrary assumption
that the interest rate and its lag are forward-looking:

\emph{Region 2, first case}: Negative unbounded output gap rule parameter
$F_{x}<0$ but where the inflation rule parameter is positive but does not
satisfy the Taylor principle is satisfied for the inflation rule parameter
$0<F_{\pi}<1$.

\emph{Region 2, second case}: Negative unbounded output gap rule parameter
$F_{x}<0$ and negative unbounded inflation rule parameter $F_{\pi}<0$.

\emph{Region 4, second case}: Negative unbounded output gap rule parameter
$F_{x}<0$ even though the Taylor principle is satisfied for unbounded large
inflation rule parameter $F_{\pi}>1.$

\emph{Region 4, third case}: Unlikely unbounded positive rule parameters
$F_{\pi}>2$, $F_{x}>1$.

These guidelines suggest contradictory and unbounded values for Taylor rule
parameters. They are hardly believable to policy-makers. Hence, New-Keynesian
theory requires additional \emph{ad hoc} boundary restrictions in order to
remain in the subset of the determinacy region with plausible parameters
$1<F_{\pi}<2$ and $0\leq F_{x}<1$. This area is similar to the condition found
in the Taylor model based on the cost of capital and negative feedback
mechanism for the inflation Taylor rule parameter: $1<F_{\pi}<1+\frac{2}{ab}$.
But the signs of the transmission mechanism on the output gap and of the
feedback mechanism are the opposite.

\subsection{Determinacy with Predetermined Interest Rate}

Let us still assume a Taylor rule. But, as an alternative thought experiment,
we can assume that, besides the two predetermined autoregressive forcing
variables, the interest rate is arbitrarily predetermined as well as its lag,
with given initial values $i_{0}$ and $i_{1}$ (or $i_{0}$ and $i_{-1}$). This
assumption is no more or no less theoretically grounded than the alternative
\emph{ad hoc} assumption of the previous section. Then the model includes four
predetermined variables. Blanchard and Kahn's (1980) determinacy condition
implies that four eigenvalues should be stable. The endogenous two-dimensional
system should be stable. This time, the determinacy area shifts to the
stability triangle with negative feedback mechanism.

Once initial values of the interest rate are given, the problem of sunspot
indeterminacy does not occur, because initial values of the forward-looking
policy target are anchored on the two values of the interest rate by the
Taylor rule (similar computations can be done for $i_{0}$ and $i_{-1}$):%

\begin{align*}
i_{0}  &  =F_{x}x_{0}+F_{\pi}\pi_{0}\\
i_{1}  &  =E_{t}i_{1}=E_{t}\left(  F_{x}x_{1}+F_{\pi}\pi_{1}\right)  =\left(
\begin{array}
[c]{cc}%
F_{x} & F_{\pi}%
\end{array}
\right)  \left(
\begin{array}
[c]{cc}%
1+\frac{\gamma\kappa}{\beta}+\gamma F_{x} & -\frac{\gamma}{\beta}+\gamma
F_{\pi}\\
-\frac{\kappa}{\beta} & \frac{1}{\beta}%
\end{array}
\right)  \left(
\begin{array}
[c]{c}%
x_{0}\\
\pi_{0}%
\end{array}
\right) \\
i_{1}  &  =\left(
\begin{array}
[c]{cc}%
F_{x}\left(  \gamma F_{x}+\frac{\kappa}{\beta}\gamma+1\right)  -\frac{\kappa
}{\beta}F_{\pi} & \frac{1}{\beta}F_{\pi}+F_{x}\left(  \gamma F_{\pi}-\frac
{1}{\beta}\gamma\right)
\end{array}
\right)  \allowbreak\left(
\begin{array}
[c]{c}%
x_{0}\\
\pi_{0}%
\end{array}
\right)  .
\end{align*}

One solves the linear system:%

\[
\left(
\begin{array}
[c]{c}%
i_{1}\\
i_{0}%
\end{array}
\right)  =\left(
\begin{array}
[c]{cc}%
F_{x}\left(  \gamma F_{x}+\frac{\kappa}{\beta}\gamma+1\right)  -\frac{\kappa
}{\beta}F_{\pi} & \frac{1}{\beta}F_{\pi}+F_{x}\left(  \gamma F_{\pi}-\frac
{1}{\beta}\gamma\right) \\
F_{x} & F_{\pi}%
\end{array}
\right)  \allowbreak\left(
\begin{array}
[c]{c}%
x_{0}\\
\pi_{0}%
\end{array}
\right)  .
\]

Initial values of forward-looking variables are anchored as follows:%

\[
\left(
\begin{array}
[c]{c}%
x_{0}\\
\pi_{0}%
\end{array}
\right)  =\left(
\begin{array}
[c]{cc}%
F_{x}\left(  \gamma F_{x}+\frac{\kappa}{\beta}\gamma+1\right)  -\frac{\kappa
}{\beta}F_{\pi} & \frac{1}{\beta}F_{\pi}+F_{x}\left(  \gamma F_{\pi}-\frac
{1}{\beta}\gamma\right) \\
F_{x} & F_{\pi}%
\end{array}
\right)  ^{-1}\left(
\begin{array}
[c]{c}%
i_{1}\\
i_{0}%
\end{array}
\right)  .
\]

They exclude sunspots although the equilibrium is a sink. The determinacy area
is the bounded stability triangle represented in Figure 1.

Consider now Figure 2, with an alternative transmission mechanism: a delayed
cost of capital and an accelerationist Phillips curve $\gamma<0$\textit{,
}$\kappa<0$\textit{.} Assume that the interest rate and its lag are
predetermined. The determinacy area is now a \emph{bounded} triangle where the
Taylor principle is satisfied for the inflation Taylor rule parameter
\textit{and} where the output gap Taylor rule parameter is positive.

By contrast, the determinacy area with the New-Keynesian transmission
mechanism $\gamma>0$\textit{, }$\kappa>0$\ assuming the interest rate and its
lag to be forward looking, includes a wide range of unbelievable
\emph{unbounded} areas, where Taylor rule parameters are too large and/or have
negative signs.

\section{RAMSEY\ OPTIMAL\ POLICY}

For the New-Keynesian transmission mechanism of section 2, Gianonni and
Woodford (2003) analyze Ramsey optimal policy in a model with an infinite time
horizon. Kara (2007) computes Ramsey optimal policy with a probability to
renege commitment in each period. The key change with respect to the infinite
horizon model is that the discount factor is now multiplied by the probability
of not reneging commitment $0<q\leq1$, so that the range of values of this
"credibility augmented discount factor", $0<\beta q\leq1$, is much larger than
for the discount factor alone: $0.98<\beta\leq1$. Several names have been
given to this kind of policy: "stochastic replanning" (Roberds, 1987) or
"quasi commitment" (Schaumburg and Tambalotti, 2007, Kara 2007). The
assumption is observationally equivalent to Chari and Kehoe's (1990) optimal
policy under sustainable plans, facing a punishment threat at a given horizon
in case of deviation from an optimal plan (Fujiwara, Kam, Sunakawa (2019)).
The discretionary equilibrium where the central bank re-optimizes continuously
without any credibility ($q=0$) is a point of measure zero with respect to the
continuous interval $q\in\left]  0,1\right]  $. Since Ramsey optimal policy
under quasi-commitment includes the case of very low credibility such as
$q=10^{-7}$ the discretionary equilibrium is not a relevant equilibrium
(Chatelain and Ralf (2018b)).

The contribution of this section is to use an intermediate step of Chatelain
and Ralf's (2019a) algorithm in order to locate the full range of reduced form
Taylor rule parameters of Ramsey optimal policy when varying the central bank
preferences in the plane of Taylor rule parameters.

The policy maker minimizes the expectation of the present value $W$ of a
discounted quadratic loss function $L_{t}$ over a finite horizon $T\geq3$. He
chooses the interest rate with respect to the policy targets (inflation and
the output gap) with a positive weight on the second policy target (output
gap) $\mu_{x}\geq0$ and a weight normalized to one, $\mu_{\pi}=1$, for
inflation (the limit case $\mu_{\pi}=0$ can also be taken into account). We
have a \emph{strictly} positive adjustment cost parameter $\mu_{i}>0$ on the
volatility of the policy instrument (the interest rate) and a discount factor
$\beta$:%

\begin{equation}
E\left(  W\right)  =-E_{t}%
{\displaystyle\sum\limits_{t=0}^{T}}
\beta^{t}\left\{  \mu_{\pi}\frac{\pi_{t}^{2}}{2}+\mu_{x}\frac{x_{t}^{2}}%
{2}+\mu_{i}\frac{i_{t}^{2}}{2}\right\}  ,T\geq3,
\end{equation}

subject to the private sector's New-Keynesian four equations model (equations
(1) to (4)), with initial conditions for the predetermined state variables and
natural boundary conditions for the private sector's forward variables.
Denoting Lagrangian multipliers $\phi_{x,t}$ for the consumption Euler
equation and $\phi_{\pi,t}$ for the New-Keynesian Phillips curve, the
Lagrangian $%
\mathcal{L}%
$ is:%

\begin{align}%
\mathcal{L}%
=-E_{0}%
{\displaystyle\sum\limits_{t=0}^{T}}
\beta^{t} \text{\{}  &  \mu_{\pi}\frac{\pi_{t}^{2}}{2}+\mu_{x}\frac{x_{t}^{2}%
}{2}+\mu_{i}\frac{i_{t}^{2}}{2}\\
+  &  \phi_{x,t}\left[  x_{t}-x_{t+1}+\sigma\left(  i_{t}-\pi_{t+1}\right)
\right]  +\phi_{\pi,t}\left[  \pi_{t}-\beta_{t}\pi_{t+1}-\kappa x_{t}\right]
\text{\}} \text{.}\nonumber
\end{align}

The law of iterated expectations has been used to eliminate the expectations
that appeared in each constraint. Because of the certainty equivalence
principle for determining optimal policy in the linear quadratic regulator
including additive normal random shocks (Simon (1956)), the expectations of
random variables $u_{t}$ are set to zero and do not appear in the Lagrangian.

Since inflation and the output gap are assumed to be forward-looking, they are
optimally chosen at the initial date $t=0$ and at the final date $t=T$
according to transversality conditions, also called natural boundary conditions:%

\begin{equation}
\phi_{\pi,t=0}=\phi_{x,t=0}=\phi_{\pi,t=T}=\phi_{x,t=T}=0\text{
(Transversality conditions).}%
\end{equation}

The Hamiltonian system with Lagrange multipliers of the linear quadratic
regulator includes two stable roots and two unstable roots:%

\begin{align}
x_{t}  &  =x_{t+1}+\sigma\left(  i_{t}-\pi_{t+1}\right) \\
\pi_{t}  &  =\beta_{t}\pi_{t+1}+\kappa x_{t}\\
\frac{\partial%
\mathcal{L}%
}{\partial\pi_{t}}  &  =0\Rightarrow\mu_{\pi}\pi_{t}+\phi_{\pi,t}-\phi
_{\pi,t-1}-\frac{\sigma}{\beta}\phi_{x,t-1}=0\text{ for }0\leq t\leq T,\\
\frac{\partial%
\mathcal{L}%
}{\partial x_{t}}  &  =0\Rightarrow\mu_{x}x_{t}-\kappa\phi_{\pi,t}+\phi
_{x,t}-\frac{1}{\beta}\phi_{x,t-1}=0\text{ for }0\leq t\leq T.
\end{align}

A bounded optimal plan is a set of bounded processes $\left\{  \pi_{t}%
,x_{t,}i_{t},\phi_{x,t},\phi_{\pi,t}\right\}  $ for date $0\leq t\leq T$ of
the Hamiltonian system that satisfy the four monetary transmission mechanism
equations, the first order equations, and the optimal initial conditions.

The link between the Lagrange multipliers and the interest rate is given by:%

\[
\frac{\partial%
\mathcal{L}%
}{\partial i_{t}}=0\Rightarrow\mu_{i}i_{t}+\sigma\phi_{x,t}=0\Rightarrow\text{
}\phi_{x,t}=-\frac{\mu_{i}}{\sigma}i_{t}\text{ for }0\leq t\leq T.
\]

The boundary conditions and the marginal conditions lead to the following
constraints on the interest rate:%
\begin{align}
\frac{\partial%
\mathcal{L}%
}{\partial i_{t}}  &  =0\Rightarrow\mu_{i}i_{t}+\sigma\phi_{x,t}%
=0\Rightarrow\text{ }\phi_{x,t}=-\frac{\mu_{i}}{\sigma}i_{t}\text{ for }0\leq
t\leq T,\\
\phi_{\pi,t=0}  &  =\phi_{x,t=0}=\phi_{\pi,t=T}=\phi_{x,t=T}=0\text{ hence }\\
i_{-1}  &  =i_{-2}=0=i_{T-1}=i_{T-2}\text{ (}i_{0}\neq0\text{ for the shortest
horizon }T=3\text{).}%
\end{align}

Using the method of undetermined coefficients, Ljungqvist and Sargent (2012,
chapter 19.3.1, \emph{step 1}, p.769) solve a Riccati equation which allows to
compute endogenous optimal negative feedback rule parameters $\left(
F_{x,R},F_{\pi,R}\right)  $. Anderson, Hansen, McGrattan and Sargent (1996,
p.203) consider that autoregressive shocks are initially set to zero
($\mathbf{z}_{0}=\mathbf{0}$) to solve this Riccati equation. They do not
depend on autoregressive parameters of exogenous variables. Anderson, Hansen,
McGrattan and Sargent (1996, p.203) take into account the exogenous
autoregressive process of forcing variables. They solve the rule parameter
responding to shocks $F_{z,R},F_{u,R}$. This step-one representation of the
Ramsey optimal policy rule depends on all four variables of the private sector:%

\begin{equation}
i_{t}=F_{x}x_{t}+F_{\pi}\pi_{t}+F_{z}z_{t}+F_{u}u_{t}.
\end{equation}

Besides the two stable eigenvalues of the block of endogenous variables, there
are the two stable eigenvalues that correspond to the autoregressive
parameters of the block of the two predetermined and exogenous forcing
variables (productivity and cost-push shock).

Ljungqvist and Sargent (2012, chapter 19, step 2) compute the linear\ stable
subspace constraint between Lagrange multipliers (in this paper: $\phi
_{x,t},\phi_{\pi,t}$) and the variables of the private sector: $\mathbf{\phi
}_{t}=\mathbf{P}_{y}\mathbf{y}_{t}+\mathbf{P}_{z}\mathbf{z}_{t}$. The matrix
$\mathbf{P}_{y}$ is the unique solution of a discrete-time Riccati equation
(Ljungqvist and Sargent (2012), chapter 19) and $\mathbf{P}_{z}$ is the unique
solution of a Sylvester equation in the augmented discounted linear quadratic
regulator (Anderson, Hansen, McGrattan and Sargent (1996), pp. 202-204).

Ljungqvist and Sargent (2012, chapter 19, step 3) substitute the
forward-looking variables (in this paper: $x_{t},\pi_{t}$) in the rule by
their Lagrange multipliers (in this paper: $\phi_{x,t},\phi_{\pi,t}$) using
the linear constraint $\mathbf{\phi}_{t}=\mathbf{P}_{y}\mathbf{y}%
_{t}+\mathbf{P}_{z}\mathbf{z}_{t}$. This representation of the Ramsey optimal
policy rule is a function of four predetermined variables ($\mathbf{\phi}%
_{t},\mathbf{z}_{t}$) which are not observable.

Ljungqvist and Sargent (2012, chapter 19, step 4) finally solve for the
optimal initial anchor of forward-looking variables on predetermined variables:%

\begin{equation}
\mathbf{\phi}_{t}=\mathbf{P}_{y}\mathbf{y}_{t}+\mathbf{P}_{z}\mathbf{z}%
_{t}\text{, }\mathbf{\phi}_{0}=\mathbf{0}\text{\textbf{, }}\mathbf{\mathbf{P}%
_{y}}\text{ invertible }\mathbf{\Longleftrightarrow y}_{0}=-\mathbf{P}%
_{y}^{-1}\mathbf{P}_{z}\mathbf{z}_{0}\text{.}%
\end{equation}

Furthermore, Ljunqgvist and Sargent (2012, chapter 19.3.7) mention another
representation of a Ramsey optimal policy rule which depends on lags of the
policy instruments and predetermined variables.

In the article of Giannoni and Woodford (2003) a different Ramsey optimal
policy rule which depends on lags of policy instruments and of forward-looking
variables (inflation and the output gap) is presented. All variables are
observable. The predetermined autoregressive forcing variables ($z_{t},u_{t}$)
are substituted by a one period lag and a two period lag of the interest rate
($i_{t-1},i_{t-2}$). Chatelain and Ralf (2018c) demonstrate that the sum of
parameters of the lags of funds rate is smaller than one for this
representation of the solution of Ramsey optimal policy, so that Giannoni and
Woodford (2003) "super-inertial" policy rules are ruled out. Other
observationally equivalent representations of the Ramsey optimal policy rule
may use leads or lags of other variables than the policy instrument.

All these various representations of Ramsey optimal policy rules correspond to
linear substitutions of the variables which are taken as the set of basis
vectors. They are all \emph{observationally equivalent} taking into account
the other equations of the stable solution of the Hamiltonian system. The
eigenvalues of the Hamiltonian system are \emph{invariant} to these changes of
basis related to a change of representation of the rule of Ramsey optimal policy.

Step 1\ representation of the rule of Ramsey optimal policy, with inflation
and output gap as rule parameters (for $\mathbf{z}_{0}=\mathbf{0}$),
corresponds to the representation of Figure 1 showing bifurcations and areas
for stable and unstable eigenvalues. For each of the infinite number of other
observationally equivalent representations of the rule of Ramsey optimal
policy, including leads or lags and/or linear changes of variables (basis
vectors) within the stable subspace of Ramsey optimal policy, a similar Figure
can be drawn to highlight stability and bifurcations.

\textbf{Proposition 7. }\emph{The Taylor rule parameters on inflation and on
the output gap in Ljunqgvist and Sargent's (2012, Chapter 19) step 1
representation are located in a subset of the stability triangle of Section 2,
which is located on the left-hand side of the center of the stability
triangle. This center corresponds to both eigenvalues equal to zero.}

\textbf{Proof. }Ramsey optimal policy assumes that inflation and the output
gap are forward-looking \emph{and that the policy instrument (the interest
rate) and its lag are predetermined.} Two stable eigenvalues in the block of
two endogenous variables of the New-Keynesian model are required to satisfy
Blanchard and Kahn's (1980) determinacy condition. The reduced form Taylor
rule parameters have to lie in the area where both eigenvalues are stable,
that is, in the stability triangle.

Q.E.D.

The key principle is that negative feedback mechanism is stabilizing
forward-looking variables, such as output and inflation, during a finite short
horizon (the duration of a monetary policy regime), leaning against inflation
and output gap spirals. 

For a numerical example, Table 2 and Figure 3\ provide the boundaries of the
triangle of the linear quadratic regulator (LQR) reduced form Taylor rule
parameters, obtained by a simulation grid, varying \emph{the weights in the
loss function in three dimensions}. The sides of the LQR triangle correspond
to the cases where the central bank minimizes only the variance of inflation
(inflation nutter) without taking into account the zero lower bound constraint
on the policy interest rate ($\mu_{i}=10^{-7}>0$), or minimizes only the
variance of output gap without taking into account the zero lower bound
($\mu_{i}=10^{-7}>0$), or seeks only maximal inertia of the policy rate
($\mu_{i}\rightarrow+\infty$).

\textbf{Table 2:} Step 1: LQR rule parameters for $\kappa=0.1$, $\gamma=0.5$,
$\beta=0.99$, $\rho_{z}=\rho_{u}=0.9$.\newline%
\begin{tabular}
[c]{|l|l|l|l|l|l|l|l|l|l|}\hline
Minimize only: & $\mu_{\pi}$ & $\mu_{x}$ & $\mu_{i}$ & $\left\vert \lambda
_{1}\right\vert $ & $\left\vert \lambda_{2}\right\vert $ & $F_{\pi}$ & $F_{x}$
& $F_{z}$ & $F_{u}$\\\hline
Inflation & $1$ & $0$ & $10^{-7}$ & $7.10^{-5}$ & $0.006$ & $21.21$ & $-3.92$
& $-2.01$ & $39.5$\\\hline
Inflation output gap & $4$ & $1$ & $10^{-7}$ & $4.10^{-7}$ & $0.819$ & $4.76$
& $-2.27$ & $-2.01$ & $17.6$\\\hline
Inflation output gap & $1$ & $1$ & $10^{-7}$ & $4.10^{-7}$ & $0.905$ & $3.03$
& $-2.10$ & $-2.01$ & $12.8$\\\hline
Inflation output gap & $1/4$ & $1$ & $10^{-7}$ & $4.10^{-7}$ & $0.951$ &
$2.10$ & $-2.01$ & $-2.01$ & $8.90$\\\hline
Output gap & $0$ & $1$ & $10^{-7}$ & $4.10^{-7}$ & $0.995$ & $1.21$ & $-1.92$
& $-2.01$ & $2.95$\\\hline
Output gap interest & $0$ & $4$ & $1$ & $0.348$ & $0.953$ & $1.70$ & $-1.31$ &
$-2.21$ & $6.78$\\\hline
Output gap interest & $0$ & $1$ & $1$ & $0.541$ & $0.918$ & $1.83$ & $-0.98$ &
$-2.43$ & $7.38$\\\hline
Output gap interest & $0$ & $1/4$ & $1$ & $0.663$ & $0.878$ & $1.87$ & $-0.82$
& $-2.42$ & $7.55$\\\hline
Interest rate & $0$ & $0$ & $1(+\infty)$ & $0.748$ & $0.833$ & $1.89$ &
$-0.74$ & $-2.46$ & $7.60$\\\hline
Inflation interest & $1/4$ & $0$ & $1$ & $0.784$ & $0.784$ & $1.99$ & $-0.77$
& $-2.43$ & $7.85$\\\hline
Inflation interest & $1$ & $0$ & $1$ & $0.772$ & $0.772$ & $2.22$ & $-0.83$ &
$-2.37$ & $8.45$\\\hline
Inflation interest & $4$ & $0$ & $1$ & $0.742$ & $0.742$ & $2.82$ & $-0.98$ &
$-2.26$ & $9.95$\\\hline
\end{tabular}
\newline

The LQR\ triangle is contained within the stability triangle, see Figure 3. A
similar analysis can be made for the alternative monetary policy transmission
mechanism with $\gamma<0$ and $\kappa<0$, see Figure 4.

An example of finding the optimal initial anchor, step 4 in the Ljungqvist and
Sargent approach is given in Table 3. Here the columns below $x_{0}$ and below
$\pi_{0}$ indicate the weights of $z_{0}$ and $u_{0}$ in the VAR.

\textbf{Table 3:} Step 4: optimal initial anchors, $\kappa=0.1,\gamma
=0.5,\beta=0.99,\rho_{z,x}=\rho_{z,\pi}=0.9$.\newline%

\begin{tabular}
[c]{|l|l|l|l|l|l|}\hline
Minimize only: & $\mu_{\pi}$ & $\mu_{x}$ & $\mu_{i}$ & $x_{0}$ & $\pi_{0}%
$\\\hline
Inflation & $1$ & $0$ & $10^{-7}$ & $10^{-4}z_{0}-10.1u_{0}$ & $10^{-5}%
z_{0}+10^{-6}u_{0}$\\\hline
Inflation output gap & $4$ & $1$ & $10^{-7}$ & $10^{-6}z_{0}-1.25u_{0}$ &
$10^{-8}z_{0}+3.14u_{0}$\\\hline
Inflation output gap & $1$ & $1$ & $10^{-7}$ & $10^{-6}z_{0}-0.49u_{0}$ &
$10^{-8}z_{0}+4.91u_{0}$\\\hline
Inflation output gap & $1/4$ & $1$ & $10^{-7}$ & $10^{-6}z_{0}-0.16u_{0}$ &
$10^{-6}z_{0}+6.66u_{0}$\\\hline
Output gap & $0$ & $1$ & $10^{-7}$ & $10^{-6}z_{0}+10^{-6}u_{0}$ &
$-10^{-6}z_{0}+9.61u_{0}$\\\hline
Output gap interest & $0$ & $4$ & $1$ & $0.35z_{0}+1.53u_{0}$ & $-0.56z_{0}%
+7.18u_{0}$\\\hline
Output gap interest & $0$ & $1$ & $1$ & $0.58z_{0}+2.52u_{0}$ & $-0.79z_{0}%
+6.20u_{0}$\\\hline
Output gap interest & $0$ & $1/4$ & $1$ & $0.72z_{0}+3.13u_{0}$ &
$-0.92z_{0}+5.63u_{0}$\\\hline
Interest rate & $0$ & $0$ & $1(+\infty)$ & $0.73z_{0}+3.14u_{0}$ &
$-0.92z_{0}+5.63u_{0}$\\\hline
Inflation interest & $1/4$ & $0$ & $1$ & $5.00z_{0}-10.3u_{0}$ &
$0.71z_{0}-0.04u_{0}$\\\hline
Inflation interest & $1$ & $0$ & $1$ & $4.45z_{0}-10.3u_{0}$ & $0.61z_{0}%
-0.03u_{0}$\\\hline
Inflation interest & $4$ & $0$ & $1$ & $3.45z_{0}-10.2u_{0}$ & $0.42z_{0}%
-0.02u_{0}$\\\hline
\end{tabular}

This leads to the following observations:

- When the central bank is an inflation nutter, both eigenvalues are close to
zero (point $\Omega$ in Figure 3 corresponds to the case when both eigenvalues
are exactly equal to zero). To stabilize inflation in the second step of the
transmission mechanism, one has to stabilize the output gap in the first step.
The optimal anchor leads to an initial jump of inflation to zero, whereas the
output gap jump is not zero.

- The lower side of the LQR\ triangle corresponds to the no-lower-bound
constraint (no cost of changing the policy rate) and to changes of the
relative weight on the output gap variance with respect to inflation variance.
The lower the weight $\mu_{\pi}$ for a given weight $\mu_{x}$, the lower
$F_{\pi}$, the response of the interest rate to inflation in the Taylor rule,
the larger the initial jump of inflation, the lower the initial jump of the
output gap.

- When the central bank is an output gap nutter, one eigenvalue is zero
(related to output gap stabilization in next period based on the Euler
consumption equation). However, the other eigenvalue tends to one when the
weight of inflation is zero. There is no margin of errors on the inflation
Taylor rule parameters with respect to the saddle-node bifurcation This second
eigenvalue is related to the New-Keynesian Phillips curve and the
autoregressive parameter of inflation, which only occurs in the second period
of the monetary policy transmission mechanism.

- The upper left side of the LQR triangle (closest to the saddle-node
bifurcation) corresponds to a weight $\mu_{\pi}=0$ for inflation and rising
relative cost of changing the policy instruments with respect to the output
gap weight. The lower the weight $\mu_{x}$ for a given weight on the policy
instrument $\mu_{i}$, the higher $F_{\pi}$ the response of interest rate to
inflation in the Taylor rule, the lower the initial jump of inflation, the
larger the initial jump of the output gap. "Mostly inflation eigenvalue"
decreases from $0.995$ to $0.833$. "Mostly output gap eigenvalue" increases
from zero to $0.748$, see Table 2.

- When the central bank uses minimum energy control (it minimizes only the
volatility of its policy instrument (or "input"): $\mu_{x}=\mu_{\pi}=0$,
$\mu_{i}>0$), the eigenvalue $0.748$ corresponds to the stable eigenvalue of
laissez-faire matrix $\sqrt{\beta}\mathbf{A}_{yy}$ and the second stable
eigenvalue $0.833=1/1.20$ is the inverse of the unstable eigenvalue $1.20$\ of
the laissez-faire matrix $\sqrt{\beta}\mathbf{A}_{yy}$.

- The upper right side of the LQR triangle corresponds to zero weight on the
output gap, and a relative increase of the weight of inflation with respect to
the weight of the policy instrument. The higher the weight $\mu_{\pi}$ for a
given weight on the policy instrument $\mu_{i}$, the inflation Taylor rule
parameter rises from $1.89$ to $21.2$ (inflation nutter case). The inflation
initial jump decreases toward zero.

\section{HOPF\ BIFURCATION}

As has been seen in the previous section, different emphasis of monetary
policy on inflation, output gap, and the interest rate lead to different
eigenvalues of the dynamic system and therefore to different stability
properties. As a result, we can state the following proposition.

\textbf{Proposition 8. }\emph{Shifting from Ramsey optimal policy (where the
interest rate and its lag are predetermined variables) to a New-Keynesian
Taylor rule with plausible Taylor rule parameters }(e.g., $1<F_{\pi}<2$
\emph{and} $0<F_{x}<1$) \emph{(where the interest rate and its lag are
forward-looking variables), when the Taylor principle is satisfied in both
regimes, corresponds to a Hopf bifurcation (crossing }$D=1$ \emph{for}
$T^{2}-4D<0$\emph{) in the New-Keynesian model.}

\textbf{Proof.}

The interest rate and its lag are optimally predetermined in a regime of
Ramsey optimal policy, whereas they are arbitrarily assumed to be
forward-looking in the determinate solution of the New-Keynesian model with a
Taylor rule.

For Ramsey optimal policy, the number of predetermined variables is equal to
four when the two predetermined forcing variables are taken into account. In
the case of determinacy in the new Keynesian model with a Taylor rule, this
number is reduced to two, namely only the two forcing variables.

This implies that the controllable part of the New-Keynesian model includes
two stable eigenvalues with Ramsey optimal policy. The policy rule parameters
lie in the stability triangle ABC bounded by $p(1)>0$, $p(-1)>0$ and $D<1$,
regions 4.1 and 4.2 in Figure 1. See also Figure 5 for a detailed view of a
smaller range of possible parameter values.

By contrast, in the New-Keynesian model we have no stable eigenvalue. The
determinacy regions are contained in regions 4.3 and 4.4 with $p(1)<0$ and
$p(-1)<0$ either with conjugate complex or real eigenvalues larger than one in
absolute values. As a plausible area for Taylor rule parameters $1<F_{\pi}<2$
and $0<F_{x}<1$ are given, see purple area in Figure 5.

Because both policy rules are in region 4 above saddle-node and flip
bifurcation limits, defined by $p(1)>0$ and $p(-1)>0$, when modelers shift
their theory on central bank behavior from Ramsey policy to New-Keynesian
determinacy with Taylor rule they move from the red area inside the stablity
triangle to the purple area outside the triangle, see Figure 5. Their change
of theory corresponds to a Hopf bifurcation, crossing the Hopf bifurcation
border $D=1$, line AB in Figures 1 and 5.

Q.E.D.

\section{CHANGING\ THE\ HYPOTHESES\ ON\ SHOCKS}

The zero variance of shocks covers two factors: the variance of the random
i.i.d. component $\varepsilon_{z,t}$ and the auto-correlation $\rho$ specific
to forcing variables. In what follows we will briefly discuss how changes of
these hypotheses affect the results.

\textbf{Zero variance of the i.i.d. components $\varepsilon_{z,t}$ and
$\varepsilon_{u,t}$ with a non-zero autocorrelation of forcing variables}

For the New-Keynesian Taylor rule, this does not change the results because
inflation and output gap can be anchored on deterministic autoregressive
forcing variables.

For Ramsey optimal policy, this does not change the results, because the
optimal endogenous policy rule parameters do not depend on the variance of
shocks in the case of the augmented linear quadratic regulator. This is the
certainty equivalent principle in the case of quadratic optimization subject
to a linear dynamic system (Simon (1956)).

\textbf{Zero autocorrelation of both forcing variables $z_{t}$ and $u_{t}$,
with a non-zero variance of the i.i.d. components $\varepsilon_{z,t}$ and
$\varepsilon_{u,t}$.}

For the New-Keynesian Taylor rule, there are two forward-looking variables
that are not predetermined. We then get a degenerate rational expectations
equilibrium. As soon as shocks $\varepsilon_{z,t}$ and $\varepsilon_{u,t}$ are
known, inflation and output gap instantaneously jump back to their respective
long-run steady-state value, which is the only anchor available. This implies
that the interest rate never deviates from its long-run steady-state value.
The Taylor rule cannot be estimated. The long run steady state for output gap
and inflation is a source for the Taylor rule parameters chosen in the
New-Keynesian determinacy area (region 4 with $D>1$ and region 2).

For Ramsey optimal policy, there are two predetermined Lagrange multipliers
set to zero, which optimally predetermine the interest rate and its lag. The
optimal anchors of inflation and the output gap at the initial date (stacked
in the vector $\mathbf{y}_{0}$) are given by $\mathbf{y}_{0}=-\mathbf{P}%
_{y}^{-1}\mathbf{P}_{z}\mathbf{z}_{0}$, where $\mathbf{z}_{0}$ corresponds to
the expectations of both autoregressive forcing variables. With zero
auto-correlation, this expectation is equal to zero for both shocks. Hence,
$\mathbf{z}_{0}=0=\mathbf{y}_{0}$. The optimal initial anchor is also the long
run steady state value of inflation and the output gap. The optimal interest
rate never deviates from its long run steady state value: this is how it is
predetermined at the initial date. The reduced form Taylor rule parameters for
inflation and the output gap are different from zero ($\mathbf{F}_{y}\neq0$).
But if both auto-correlation coefficients are equal to zero, $\rho
_{i}=0\Rightarrow\mathbf{F}_{z}=0$, the interest rate does not respond to
non-autoregressive shocks. This is consistent with Simon's (1956) certainty
equivalence principle. The long run steady state for output gap and inflation
is a sink for the Taylor rule parameters chosen in the Ramsey optimal policy
determinacy area (region 4 with $D<1$). This equilibrium corresponds to
degenerate Ramsey optimal policy.

\textbf{Zero auto-correlation of only one of the forcing variables ($u_{t}$).}

For the New-Keynesian Taylor rule, there are two forward-looking forcing
variables for one predetermined forcing variable. For all periods, both,
inflation and the output gap, are linear functions of $z_{t}$: $\pi_{t}%
=N_{\pi}z_{t}$ and $x_{t}=N_{x}z_{t}$. Hence, the output gap is a linear
function of inflation $x_{t}=N_{x}N_{\pi}^{-1}\pi_{t}$. The recursive dynamic
system boils down to a dynamic system of dimension one. The feedback rule
needs to respond to only one linear combination of the two policy targets.
This implies selecting one linear identification restriction for the Taylor
rule parameters among an infinity of possibilities: $F_{x}=\alpha F_{\pi}$
with $\alpha\in%
\mathbb{R}
$ or $F_{\pi}=0$. For $F_{\pi}=0$ and $F_{x}\neq0$, the Taylor rule responds
only to the output gap ($i_{t}=F_{x}x_{t}$) which seemingly contradicts
inflation targeting. It is observationally equivalent to another policy rule
$i_{t}=F_{x}N_{x}N_{\pi}^{-1}\pi_{t}$, where we define $F_{\pi,2}=F_{x}%
N_{x}N_{\pi}^{-1}\neq0$ and $F_{x,2}=0$, which describes "inflation targeting"
and seemingly contradicts an output gap stabilization objective. Assuming\ a
unique predetermined variable for two forward-looking variables forces the
coincidence that both forward-looking variables are exactly collinear. This
assumption cannot "demonstrate" inflation targeting and deny an output gap
stabilization objective, because they are exactly equivalent. The determinacy
area for Taylor rule parameters is unchanged (region 4 with $D>1$ and region
2), but the chosen identification restriction has to be taken into account .

For Ramsey optimal policy, additional to the two forward-looking variables for
one predetermined forcing variable there are two backward-looking Lagrange
multipliers on inflation and on the output gap set to zero at the initial
date. \emph{At the date of the initial optimization}, both, initial inflation
and initial output gap, are anchored as linear functions to the initial value
of the unique forcing variable $z_{0}$, according to the formula:
$\mathbf{y}_{0}=-\mathbf{P}_{y}^{-1}P_{z}z_{0}$. Hence, at the initial date,
the output gap is a linear function of inflation. But, during all the periods
where the central bank does not re-optimize, the recursive dynamic system
boils down to a dynamic system of dimension three. The feedback rule responds
to output gap and inflation with unchanged formula for $\mathbf{F}_{y}$. The
determinacy area for Ramsey optimal policy is unchanged (region 4 with $D<1$).
Since the auto-correlation coefficient is now zero, the feedback rule responds
only to the shock $z_{t}$ with a parameter $F_{z}$ found by solving a scalar
Sylvester equation. The identification restriction is then $F_{u}=0$, which is
implied by $\rho_{u}=0$ in a Sylvester equation of the augmented linear
quadratic regulator.

\section{CONCLUSION}

In this paper we derived the dynamic properties of New-Keynesian Taylor rule
policy and Ramsey optimal policy. Stability, determinacy, and anchors of
policy variables were compared. According to the choice of the policy
parameters, namely, the Taylor rule parameters corresponding to the inflation
and the output target, the steady state can be a source or a sink with real or
complex eigenvalues. Changing the policy doctrine and as a consequence the
policy parameters, may change the stability properties, and bifurcations are
possible. Moving in particular from Ramsey optimal policy to a New-Keynesian
Taylor rule can be interpreted as a Hopf bifurcation.

Ramsey optimal policy can also take into account monetary and fiscal policy
(Cardani et al. (2018), Gomis-Porqueras and Zhang (2019), Chatelain and Ralf
(2019d, 2019e)). Further research may test these models, following Chatelain
and Ralf (2017) tests and estimations in the case where the transmission
mechanism is only the New-Keynesian Phillips curve. A key insight is that
positive feedback stabilization policy requires more structural parameters to
fit the data than negative feedback stabilization policy (Chatelain and\ Ralf
(2018a)). This may lead to parameter identification issue (Chatelain and Ralf (2014)).

\section*{APPENDIX A: Scilab code}

Download the open source software Scilab and copy and paste the following code
in the command window, for given preferences of the central bank (Qpi, Qx, R)
and given monetary policy transmission mechanism parameters (beta1, gamma1,
kappa, rho1, rho2). Transition matrix is multiplied by $\sqrt{\beta}$ in order
to take into account discounting as proposed by Anderson, Hansen, McGrattan
and Sargent (1996). Formulas for Riccati, Sylvester and rule parameters are
taken in Anderson, Hansen, McGrattan and Sargent (1996).

Qpi=4;Qx=0;R=1;Qxpi=0;

beta1=0.99; gamma1=0.5; kappa=0.1;

rho1=0.9; rho2=0.9; rho12=0;

Qxrho1=0; Qpirho1=0;

Qxrho2=0; Qpirho2 =0;

A1=[1-(kappa*gamma1/beta1) -gamma1/beta1 ; -kappa/beta1 1/beta1] ;

A=sqrt(beta1)*A1;

B1=[gamma1 ; 0];

B=sqrt(beta1)*B1;

Q=[Qx Qxpi ;Qxpi Qpi ];

Big=sysdiag(Q,R);

[w,wp]=fullrf(Big);

C1=wp(:,1:2);

D12=wp(:,3:\$);

M=syslin('d',A,B,C1,D12);

[Fy,Py]=lqr(M);

A+B*Fy;

Py

Fy

spec(A+B*Fy)

abs(spec(A+B*Fy))

A+B*Fy

A

B

Ayz=[-1 gamma1/beta1 ; 0 -1/beta1 ];

Azz=[rho1 rho12; rho12 rho2 ];

Qyz=[Qxrho1 Qpirho1 ; Qxrho2 Qpirho2 ];

BS=-Azz;

AS=(A+B*Fy)';

CS=Qyz+AS*Py*Ayz;

Pz=sylv(AS, BS, CS, 'd');

AS*Pz*BS+Pz-CS;

norm (AS*Pz*BS+Pz-CS);

N=-inv(Py)*Pz;

Fz=inv(R+B'*Py*B)*B'*(Py*Ayz + Pz*Azz);

sp1=spec(A+B*Fy)

sp1t=sp1'

Py

Pz

Spectrum=[sp1t rho1 rho2 ]

abs(spec(A+B*Fy))

F=[Fy Fz ]

N

\section*{APPENDIX B: Wonham theorem and pole placement}

Let $\mathbf{A}$ and $\mathbf{B}$ be real matrices of dimension $n\times n$
and $n\times m$ respectively. Let $\Lambda=\left\{  \lambda_{1},...,\lambda
_{n}\right\}  $ be an arbitrary set of $n$ complex numbers $\lambda_{i}$ such
that any $\lambda_{i}$ with Im$\left(  \lambda_{i}\right)  \neq0$ appears in
$\Lambda$ in a conjugate pair.

\textbf{Wonham (1967) pole placement theorem.} \textit{The pair }$\left(
\mathbf{A},\mathbf{B}\right)  $\textit{ is controllable (}$rank\left(
\mathbf{B,AB},\mathbf{A}^{2}\mathbf{B,}...,\mathbf{A}^{n-1}\mathbf{B}\right)
=n$\textit{) if and only if, for every choice of the set }$\Lambda=\left\{
\lambda_{1},...,\lambda_{n}\right\}  $\textit{, there is a matrix }%
$\mathbf{F}$\textit{ such that }$\mathbf{A+BF}$\textit{ has }$\Lambda$\textit{
for its sets of eigenvalues.}

Wonham's (1967) pole placement theorem states that linear feedback rule
parameters are always bifurcation parameters of controllable linear systems.
Close to bifurcations limit values, a small change of the policy rule
parameters leads to big qualitative change of the dynamics of the system to be
controlled, as eigenvalues shift from being outside the unit circle to inside
the unit circle.

\textbf{Pole placement using the canonical form.}

The characteristic polynomial for the open-loop eigenvalues and the desired
closed-loop eigenvalues corresponds to distinct coefficients (the trace and determinant):%

\[
\lambda^{2}-T\left(  \mathbf{A}\right)  \lambda+D\left(  \mathbf{A}\right)
=0\text{ and }\lambda^{2}-T\left(  \mathbf{A+BF}\right)  \lambda+D\left(
\mathbf{A+BF}\right)  =0.
\]

The canonical form of the dynamic system is such that:%

\[
\widehat{A}=\left(
\begin{array}
[c]{cc}%
T\left(  \mathbf{A}\right)  & -D\left(  \mathbf{A}\right) \\
1 & 0
\end{array}
\right)  \text{, }\widehat{B}=\left(
\begin{array}
[c]{c}%
1\\
0
\end{array}
\right)  .
\]

Hence, the closed-loop canonical model has the property that the feedback rule
parameters appear only on the first line. It should correspond to given
closed-loop trace and determinant.%

\[
\widehat{A}+\widehat{B}\widehat{F}=\left(
\begin{array}
[c]{cc}%
T\left(  \mathbf{A}\right)  +\widehat{F}_{x} & -D\left(  \mathbf{A}\right)
+\widehat{F}_{\pi}\\
1 & 0
\end{array}
\right)  =\left(
\begin{array}
[c]{cc}%
T\left(  \mathbf{A+BF}\right)  & -D\left(  \mathbf{A+BF}\right) \\
1 & 0
\end{array}
\right)  .
\]

Then:%

\[
\widehat{F}=\left(
\begin{array}
[c]{cc}%
\widehat{F}_{x} & \widehat{F}_{\pi}%
\end{array}
\right)  = \left(
\begin{array}
[c]{cc}%
T\left(  \mathbf{A+BF}\right)  -T\left(  \mathbf{A}\right)  & -\left(
D\left(  \mathbf{A+BF}\right)  -D\left(  \mathbf{A}\right)  \right)
\end{array}
\right)  .
\]

The feedback rule is equal to $\widehat{\mathbf{F}}$ times a similarity matrix
which is the product of the canonical controllability matrix of
$\widehat{\mathbf{A}}\mathbf{+}\widehat{\mathbf{B}}\widehat{\mathbf{F}}$ and
the inverse of the controllability matrix of $\mathbf{A+BF}$:%

\begin{align*}
\mathbf{F}  &  =-\left(
\begin{array}
[c]{cc}%
T\left(  \mathbf{A}\right)  -T & D-D\left(  \mathbf{A}\right)
\end{array}
\right)  \text{ }\left(
\begin{array}
[c]{cc}%
1 & T\left(  \mathbf{A}\right) \\
0 & 1
\end{array}
\right)  \left(
\begin{array}
[c]{cc}%
\gamma & \gamma\left(  T\left(  \mathbf{A}\right)  -\frac{1}{\beta}\right) \\
0 & -\frac{\kappa\gamma}{\beta}%
\end{array}
\right)  ^{-1}\\
\mathbf{F}  &  =-\left(
\begin{array}
[c]{cc}%
T\left(  \mathbf{A}\right)  -T\left(  \mathbf{A+BF}\right)  & D\left(
\mathbf{A+BF}\right)  -D\left(  \mathbf{A}\right)
\end{array}
\right)  \left(
\begin{array}
[c]{cc}%
\frac{1}{\gamma} & -\frac{1}{\kappa\gamma}\\
0 & -\frac{\beta}{\kappa\gamma}%
\end{array}
\right) \\
\mathbf{F}  &  \mathbf{=}\left(
\begin{array}
[c]{cc}%
\frac{1}{\gamma}\left(  T\left(  \mathbf{A+BF}\right)  -T\left(
\mathbf{A}\right)  \right)  & -\frac{1}{\gamma\kappa}\left(  T\left(
\mathbf{A+BF}\right)  -T\left(  \mathbf{A}\right)  \right)  +\frac{\beta
}{\gamma\kappa}\left(  D\left(  \mathbf{A+BF}\right)  -D\left(  \mathbf{A}%
\right)  \right)
\end{array}
\right)  .
\end{align*}

\textbf{Pole placement using Ackermann's (1972) formula} for a controllable
system (we use the notation $\mathbf{A+BF}$ instead of $\mathbf{A-BF}$, hence
the minus sign at the beginning):%

\[
\mathbf{F=-}\left(
\begin{array}
[c]{cc}%
0 & 1
\end{array}
\right)  \left(
\begin{array}
[c]{cc}%
\mathbf{B} & \mathbf{AB}%
\end{array}
\right)  ^{-1}\left(  \mathbf{A}^{2}-T\left(  \mathbf{A+BF}\right)
\mathbf{A}+D\left(  \mathbf{A+BF}\right)  \right)  .
\]

Applied on the New-Keynesian model:%

\begin{align*}
\mathbf{F}  &  \mathbf{=-}\left(
\begin{array}
[c]{cc}%
0 & 1
\end{array}
\right)  \left(
\begin{array}
[c]{cc}%
\gamma & \gamma\left(  \frac{\kappa}{\beta}\gamma+1\right) \\
0 & -\frac{\kappa}{\beta}\gamma
\end{array}
\right)  ^{-1}\left(  \left(
\begin{array}
[c]{cc}%
1+\frac{\gamma\kappa}{\beta} & -\frac{\gamma}{\beta}\\
-\frac{\kappa}{\beta} & \frac{1}{\beta}%
\end{array}
\right)  ^{2}-T\left(
\begin{array}
[c]{cc}%
1+\frac{\gamma\kappa}{\beta} & -\frac{\gamma}{\beta}\\
-\frac{\kappa}{\beta} & \frac{1}{\beta}%
\end{array}
\right)  +D\left(
\begin{array}
[c]{cc}%
1 & 0\\
0 & 1
\end{array}
\right)  \right) \\
\mathbf{F}  &  \mathbf{=}\left(
\begin{array}
[c]{cc}%
\frac{1}{\gamma}\left(  T\left(  \mathbf{A+BF}\right)  -T\left(
\mathbf{A}\right)  \right)  & -\frac{1}{\gamma\kappa}\left(  T\left(
\mathbf{A+BF}\right)  -T\left(  \mathbf{A}\right)  \right)  +\frac{\beta
}{\gamma\kappa}\left(  D\left(  \mathbf{A+BF}\right)  -D\left(  \mathbf{A}%
\right)  \right)
\end{array}
\right)  .
\end{align*}

The transfer function of the New-Keynesian model is:%

\[
C\left(  sI-A\right)  ^{-1}B=-\left(
\begin{array}
[c]{cc}%
1 & 1
\end{array}
\right)  \left(
\begin{array}
[c]{cc}%
1+\frac{\gamma\kappa}{\beta}-s & -\frac{\gamma}{\beta}\\
-\frac{\kappa}{\beta} & \frac{1}{\beta}-s
\end{array}
\right)  ^{-1}\left(
\begin{array}
[c]{c}%
\gamma\\
0
\end{array}
\right)  =\frac{\gamma s-\frac{\gamma+\gamma\kappa}{\beta}}{s^{2}-T_{A}%
s+D_{A}}.
\]

\end{document}